\documentclass[sigconf]{acmart}

\usepackage{amsmath,amsfonts} %
\usepackage{graphicx}%
\usepackage{textcomp}%
\usepackage{xcolor}%
\usepackage{array}%
\usepackage{pifont}%
\usepackage{amsthm}%
\usepackage{bm}
\usepackage[linesnumbered,ruled,vlined]{algorithm2e}%
\usepackage{setspace}%
\theoremstyle{remark}

\usepackage{multirow}%
\usepackage{siunitx}
\usepackage{footnote}%
\usepackage[utf8]{inputenc}
\usepackage[english]{babel}
\usepackage{blindtext}
\usepackage{dblfloatfix}%
\usepackage{xcolor,colortbl} 
\usepackage{url}

\usepackage{breakurl}
\usepackage{hyperref}
\usepackage{ctable} 
\graphicspath{ {pdf/} }
\usepackage{pifont}

\usepackage{filecontents}
\usepackage{float}

\newlength{\bibitemsep}
\newlength{\bibparskip}
\let\oldthebibliography\thebibliography
\renewcommand\thebibliography[1]{%
	\oldthebibliography{#1}%
	\setlength{\parskip}{\bibitemsep}%
	\setlength{\itemsep}{\bibparskip}%
}

\AtBeginDocument{%
  \providecommand\BibTeX{{%
    \normalfont B\kern-0.5em{\scshape i\kern-0.25em b}\kern-0.8em\TeX}}}

\copyrightyear{2022}
\acmYear{2022}
\setcopyright{rightsretained}
\acmConference[ACSAC '22]{Annual Computer Security Applications Conference}{December 5--9, 2022}{Austin, TX, USA}
\acmBooktitle{Annual Computer Security Applications Conference (ACSAC
'22), December 5--9, 2022, Austin, TX, USA}
\acmDOI{10.1145/3564625.3564638}
\acmISBN{978-1-4503-9759-9/22/12}

\begin{document}

\title{
BayesImposter: Bayesian Estimation Based .bss Imposter Attack on Industrial Control Systems
}



\author{Anomadarshi Barua, Lelin Pan, and Mohammad Abdullah Al Faruque}
\affiliation{%
  \institution{Department of Electrical Engineering and Computer Science}
  \city{University of California, Irvine}
  \state{California}
  \country{USA.}\\
  Email: \textit{\{anomadab, lelinp, alfaruqu\}@uci.edu}
}


\begin{abstract}

Over the last six years, several papers used  memory deduplication  to trigger various security issues, such as leaking heap-address and causing bit-flip in the physical memory. The most essential requirement for successful memory deduplication is to provide identical copies of a physical  page. Recent works use a brute-force approach to create identical copies of a physical  page that is an inaccurate and time-consuming primitive from the attacker's perspective.

Our work begins to fill this gap by providing a domain-specific structured way to duplicate a physical page in cloud settings in the context of industrial control systems (ICSs). Here, we show a new attack primitive - \textit{BayesImposter}, which points out that the attacker can duplicate the .bss section of the target control DLL file of cloud protocols using the \textit{Bayesian estimation} technique. Our approach results in less memory (i.e., 4 KB compared to GB) and time (i.e., 13 minutes compared to hours) compared to the brute-force approach used in recent works. We point out that ICSs can be expressed as state-space models; hence, the \textit{Bayesian estimation} is an ideal choice to be combined with memory deduplication for a successful attack in cloud settings. To demonstrate the strength of \textit{BayesImposter}, we create a real-world automation platform using a scaled-down automated high-bay warehouse and industrial-grade SIMATIC S7-1500 PLC from Siemens as a target ICS. We demonstrate that \textit{BayesImposter} can predictively inject false commands into the PLC that can cause possible equipment
damage with machine failure in the target ICS. Moreover, we show that \textit{BayesImposter} is capable of adversarial control over the target ICS resulting in severe consequences, such as killing a person but making it looks like an accident. Therefore, we also provide countermeasures to prevent the attack.

\end{abstract}

\ccsdesc[500]{Security and privacy~Embedded systems security}

\keywords{PLCs, DLL file, bayesian estimation, adversarial control}

\maketitle

\vspace{-0.0em}
\section{Introduction}

Historically, Industrial Control Systems (ICSs) follow the ANSI/ISA 95 model \cite{scholten2007road}, where \textit{disconnected} computer systems and \textit{isolated} sensor frameworks were used to screen various operations and tasks in lower \textit{levels} of the \textit{automation pyramid} \cite{bartodziej2017concept}. As we enter the fourth industrial revolution \cite{lasi2014industry} (Industry 4.0), the ANSI/ISA95 model 
is going under different transformations. These transformations include the vertically/horizontally \textit{interconnected} and \textit{decentralized} ICSs in all levels of the \textit{automation pyramid} for flexible monitoring and control. The decentralization of ICSs in Industry 4.0 adds fuel to movement to the Industrial Internet of Things (IIoT) trend, where \textit{cloud servers} and \textit{virtualization} \cite{xing2012virtualization} play an important role by providing easy-to-access automation platforms. 

In Industry 4.0, Infrastructure-as-a-Service (IaaS) enables Programmable Logic Controllers (PLCs) to connect with clouds \cite{langmann2016plc}. Moreover, to support multiple PLCs and supervisory platforms, today's ICSs use  multiple Virtual Private Servers (VPSs) in a single cloud platform \cite{givehchi2014control}. The cloud server has memory deduplication feature enabled \cite{deng2017memory}, which is a \textit{widespread optimizing} feature present in today's cloud servers to support virtualization. In this typical ICS platform, the user sends control programming and 
supervisory commands from VPSs using cloud protocols (i.e., MQTT, AMQP) to PLCs \cite{langmann2019plc}. The cloud protocol's software stack has a specific DLL file, which  transports these commands and is located in the server computer. We call this specific DLL file as \textit{target control DLL} file. 



In this paper, at first, we show that \textit{the .bss section} of the target control DLL file of cloud protocols transports the critical control commands from VPSs to PLCs (i.e., lower level of the automation pyramid). Next, after identifying the target control DLL file,  we introduce the \textit{Bayesian estimation} by which an attacker can recreate or fake the memory page of the .bss section of the target control DLL file. We name the fake .bss section\footnotemark as the \textit{.bss imposter} and denote the attack model by \textit{BayesImposter}.

The intuition behind \textit{BayesImposter} is that as ICSs can be expressed as state-space models \cite{friedland2012control}, our \textit{BayesImposter} exploits the \textit{Bayesian estimation} technique to accurately predict the current state of the industrial controller. As control commands are directly related to the current states of the industrial controller, after estimating the states, the attacker can also estimate the control commands from the estimated states. As the .bss section contains the control commands, hence, the attacker can successfully recreate the .bss section using the estimated  control commands. We show that our proposed  \textit{Bayesian estimation} results in less memory and attack time  to recreate the page of the \textit{.bss imposter} compared to the brute force approach demonstrated in recent works  \cite{razavi2016flip, barresi2015cain, bosman2016dedup,oliverio2017secure}. 


\footnotetext{In this paper, \textit{the .bss section} means the .bss section of the target control DLL file of cloud protocols; unless otherwise mentioned.}

After recreating the fake .bss section, \textit{BayesImposter} uses the underlying memory deduplication feature enabled in the cloud to merge the page of the fake .bss section with the legitimate .bss section. In this way, the attacker can locate the memory address of the fake .bss section in the host machine and can use a malicious  \textit{co-located  VPS} to trigger a bit-flip in the page of the .bss section using the Rowhammer bug \cite{razavi2016flip, barresi2015cain, bosman2016dedup,oliverio2017secure} of the host machine. 
As the .bss section contains the control commands, this paper shows that a bit flip in this section may cause corruption or even change the actual command. This method can be termed as \textit{false command injection}. The injected false commands propagate from VPSs to the PLCs and may cause an unplanned behavior with catastrophic machine failure in the target ICS. It is worthwhile to mention here that, as \textit{BayesImposter} has more control over the recreation of a fake .bss section, our attack is capable of \textit{adversarial control} over the target ICS from a \textit{co-located  VPS} on the same cloud. To the best of our knowledge,  \textit{BayesImposter} is the first work that successfully merges the idea of \textit{Bayesian estimation} of the state-space models of ICSs with the memory deduplication and the Rowhammer bug in cloud settings in the context of ICSs. 

\vspace{0.0em}
\noindent \textbf{Technical Contributions}: Our contributions are:

  
    $\bullet$ We are the first to point out how the .bss section of the target control DLL file of cloud protocols can be exploited by using  memory deduplication  in modern ICSs. 
  
  $\bullet$ We are the first to introduce Bayesian estimation to recreate the .bss section. Our attack requires less memory and time  compared to the brute force approach used in recent works \cite{razavi2016flip, barresi2015cain, bosman2016dedup,oliverio2017secure}.
  
  $\bullet$ We create a real-world scaled-down factory model of a practical ICS, which has an automated high-bay warehouse from \textit{fischertechnik} \cite{FactorySimulation}. We use an industrial-grade PLC with a  part\# SIMATIC S7-1500 \cite{SIMATIC} from Siemens to create the automation platform and connect the PLC to clouds using industry-standard cloud protocols. 
  
   $\bullet$ We evaluate \textit{BayesImposter} in our factory model considering five variants of industry-standard cloud protocols and show the adversarial control to generalize our attack model in cloud settings. The demonstration of our work is shown in the following link: {\color{blue}\url{https://sites.google.com/view/bayesmem/home}}.
  
  

\vspace{-.10em}
\section{Background}
\label{sec:background}


\vspace{-0.0em}
\subsection{\textbf{Connecting PLCs with clouds}}
\label{subsec:PLCs and clouds}


IIoT enables PLCs to upload the acquired data directly to clouds \cite{sajid2016cloud}. PLCs are connected to clouds normally in  two ways: using an adapter or directly using a standard protocol. Standard cloud protocols, such as MQTT and AMQP support \textit{bidirectional} and \textit{event-based} data transmission between PLCs and upper managements. The upper management  can modify control functions of PLCs in run-time by flashing new \textit{control programs} to PLCs from clouds.

\begin{figure}[ht!]
\vspace{-1.20em}
\centering
\includegraphics[width=0.45\textwidth,height=0.15\textheight]{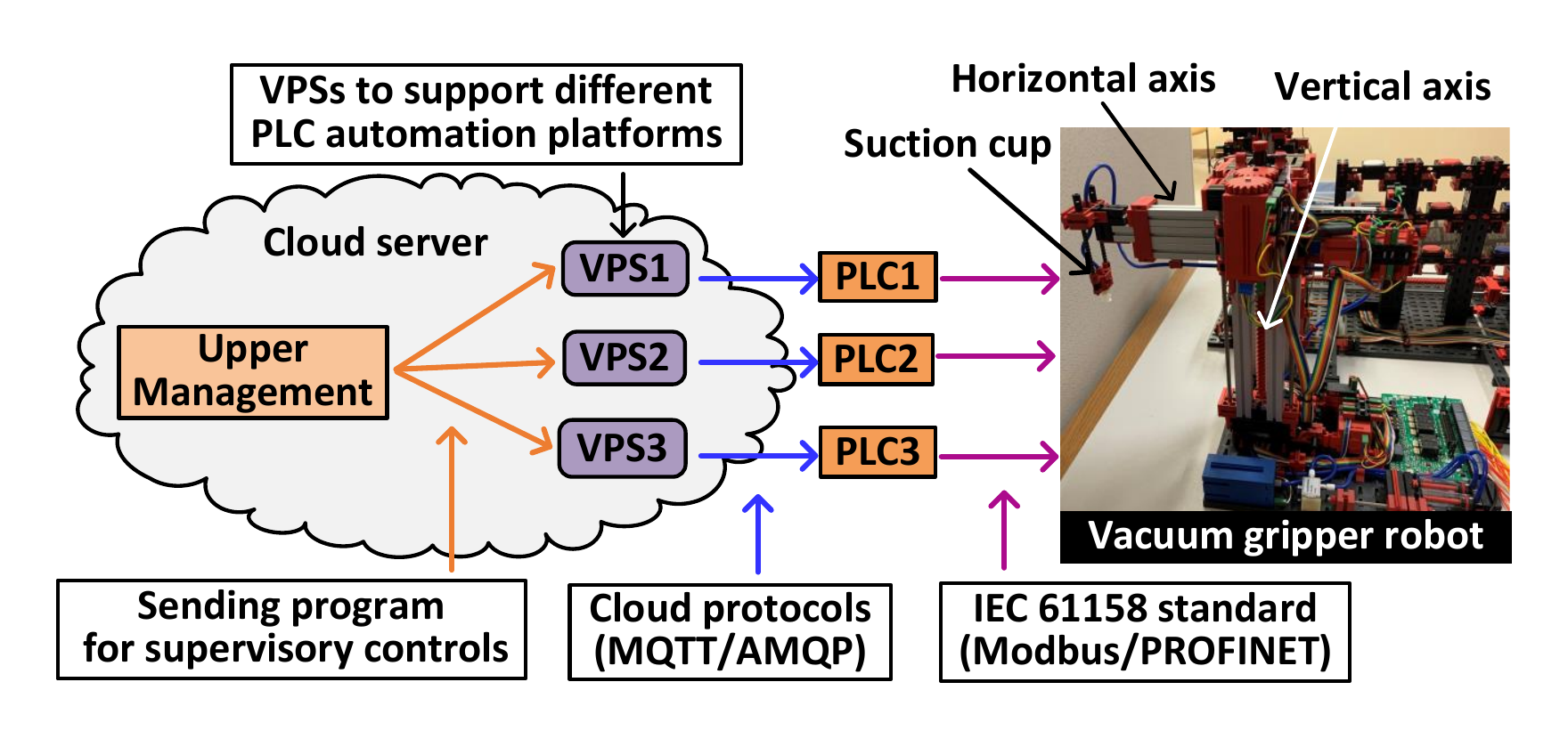}
\vspace{-1.85em}
\caption{Different components of an ICS in cloud settings.}
\label{fig:synergy_ics}
\vspace{-1.00em}
\end{figure}

\vspace{-0.95em}
\subsection{\textbf{Programs for supervisory controls}}
\label{subsec:Control programming of PLCs}

The IEC 61131 programming standard \cite{tiegelkamp1995iec} is used for control programming of PLCs. Control programs can be broadly divided into three categories: (i) programs for basic functions, (ii) programs for supervisory controls, and (iii) programs for critical time-constraint functions (e.g., security and real-time response, etc.). Traditionally, all these three categories of  control programs were implemented in PLCs in industrial premises. However, with the new trend in Industry 4.0, nowadays, only the programs for critical time-constraint functions are implemented in PLCs. Programs for basic functions and supervisory controls are not implemented in PLCs; rather, they are implemented in clouds or in web-server. For example, basic functions and supervisory control programs are outsourced as web services to a cloud or to a server for class C33 PLC controller \cite{langmann2019plc}. \textit{This gives more flexibility to upper managements as they can change programs remotely in run-time to tackle abruptly changing situations.} 

\vspace{-0.3em}
\subsection{\textbf{Use of VPSs with PLCs}}
\label{subsec:PLCs and Virtual Machines}

ICSs are becoming more complex in Industry 4.0. ICSs often need to support multiple automation platforms that may conflict with each other. Moreover, multiple PLC controllers and supervisory platforms may need multiple software packages that may require multiple operating systems. Also, introducing web servers and clouds to ICSs increases the necessity of using multiple private servers. As using multiple separate physical machines to support multiple automation platforms or operating systems or private servers is one of the available solutions, industries evidently use VPSs to reduce the number of required physical machines to reduce  cost \cite{ruotsalainen2018hardening}. Moreover, modern cloud platforms offer cheap access to VPSs by sharing  a single server among multiple operating systems on a single server machine using virtualization software \cite{siemensVPS}. 


\vspace{-0.0em}
\begin{figure*}[ht!]
\centering
\includegraphics[width=0.85\textwidth,height=0.16\textheight]{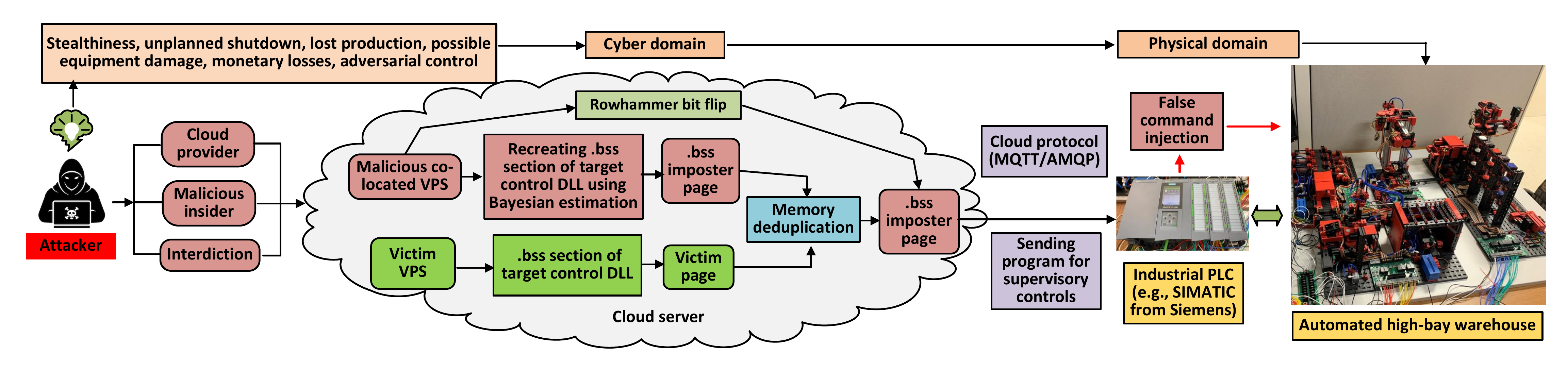}
\vspace{-1.8em}
\caption{Different components of our attack model - \textit{BayesImposter} on industrial control systems in cloud settings.}
\label{fig:attack_model}
\vspace{-0.1100em}
\end{figure*}

\vspace{-0.500em}
\subsection{\textbf{A motivational example of an ICS}}
\label{subsec:An example of an ICS}

A motivational example is shown in Fig. \ref{fig:synergy_ics} where we consider an automated high-bay warehouse as our example ICS. It has a vacuum gripper robot, which stores objects in the storage rack of the warehouse using a suction cup and moves along the horizontal and vertical axis. We elaborate more on this in Section \ref{sec:Modelling a typical cooling system} while demonstrating our attack model. Here, multiple PLCs having different platforms are supported by a cloud using multiple VPSs. Upper management located in the cloud send programs for supervisory controls from VPSs to PLCs using cloud protocols (i.e., MQTT/AMQP). PLCs communicate with the underlying sensors and controllers using IEC 61158 standard protocols (e.g., Modbus, PROFINET, etc.).  Given this background, an attacker can perturb the supervisory control commands (i.e., false command injection) in our example ICS  and remotely hamper its normal operation using our attack model - \textit{BayesImposter}. 

\vspace{-0.500em}
\subsection{\textbf{Memory deduplication}}
\label{subsec:Memory deduplication}


Memory deduplication  is a process that merges identical pages in the physical memory into one page to reduce redundant pages having similar contents. It is a widely used feature in cloud servers allowing multiple VPSs to run on less allocated memory  in a single physical machine. The amount of redundant pages can be as high as 86\% \cite{chang2011empirical} and memory deduplication can save up to 50\% of the allocated memory in the cloud server \cite{gupta2010difference}. This feature is available in Windows 8.1, Windows Server 2016, 2019, and 2022 and Linux distribution.
Windows Servers have it as Data Deduplication \cite{datadedup} and Linux distributions have it as  Kernel Samepage Merging (KSM), which is implemented in Kernel-based Virtual Machine (KVM) 
(see Appendix \ref{subsubsec:Memory deduplication}, \ref{subsubsec:Use of Kernel Samepage Merging}, and \ref{subsec:KSM data structure} for more detail on this topic).

\vspace{-0.50em}
\section {Attack model}
\label{sec:attack model}

Fig. \ref{fig:attack_model} shows the attack model - \textit{BayesImposter} in cloud settings. The essential components of \textit{BayesImposter} are described below.

\vspace{0.0em}

\vspace{0.1em}
\textbf{(i) Target system:} We consider an infrastructure \cite{goldschmidt2015cloud} where PLCs are connected with a cloud for maintenance and control programming, and multiple Virtual Machines (VMs) acting as VPSs are located in the same cloud to support multiple automation platforms. 
As multiple VPSs in the same cloud share the same hardware, an attacker can exploit the shared hardware from a co-located VPS. 

\vspace{0.0em}
\textbf{(ii) Attacker's capabilities:} Let us consider a scenario where a user gives commands from his proprietary VPS 
to a  PLC to do control programming and supervisory controls. 

\textbf{ $\bullet$ .bss imposter:} A few specific DLL files (i.e., target control DLL) of the cloud protocols transport these commands from VPS to PLCs. These DLL files are organized into different \textit{sections}. Each section can be writable or read-only and can encapsulate executable (i.e., code) or non-executable (i.e., data) information. The section, which encapsulates uninitialized data, is known as \textit{.bss} section. The .bss section of the target control DLL contains control programming and supervisory control specific information/data, which are mostly \textit{boolean} type coming from the user as commands. 
This .bss section  is page-aligned in virtual memory as well as in physical memory. Let us denote this as \textit{victim page}. If an attacker can recreate the \textit{victim page}, the attacker can use this \textit{recreated victim page} (a.k.a., \textit{.bss imposter page}) to trigger memory deduplication.

\textbf{ $\bullet$ Bottleneck:} To recreate the \textit{victim page}, the attacker needs to guess all the initialization values of uninitialized variables of the .bss section. As there could be hundreds of control variables present in the .bss section, this is \textbf{almost impossible} for the attacker to successfully guess the \textit{victim page} and recreate it following the brute force approach adopted in recent works \cite{razavi2016flip, barresi2015cain, bosman2016dedup,oliverio2017secure}. The brute force approach was successful in \cite{razavi2016flip, barresi2015cain, bosman2016dedup,oliverio2017secure} because they only guessed a specific 32-bit data to recreate a \textit{victim page}. To guess hundreds of variables in the .bss section, the brute force approach could require hundreds of hours. Moreover, the attacker may need to  spray the physical memory with terabyte amount of recreated pages to initiate a successful attack in the brute-force approach.

\textbf{ $\bullet$ Solution:} Thankfully this challenge can be handled by using \textit{BayesImposter}. The intuition behind \textit{BayesImposter} is that  if  an  attacker  knows  the state-space  model  of  the  ICS,  the  attacker  can  estimate  the boolean and non-boolean control  commands because the control commands are directly correlated with the current states of an ICS.  As  the  .bss  section transports the control commands, the estimation of the control commands helps the attacker to successfully guess the control variables present in the .bss section leading to a successful recreation of the \textit{victim page} (i.e., \textit{.bss imposter page}).

\textbf{ $\bullet$ Memory deduplication + Rowhammer:} After recreating the .bss imposter page using our \textit{BayesImposter}, the attacker can initiate memory deduplication to merge the \textit{victim page} with the attacker's provided \textit{.bss imposter page}. In this way, the attacker maps the \textit{victim page} in his address space to initiate the Rowhammer attack on the \textit{.bss imposter page} from his address space. It can flip bits in the \textit{.bss imposter page} and change values of control commands.


\vspace{0.0em}
\textbf{(iii) Outcomes of the attack:}
As the .bss section contains important data dedicated to control programming and supervisory controls, the bit flips in the .bss section may lead to potential failure in ICSs. It can cause an unplanned shutdown, possible equipment damage, catastrophic machine failure, monetary losses, or even can kill a person but making it looks like an accident in the target ICS. 

\vspace{0.0em}
\textbf{(iv) Attacker's access level:} Our attack requires the deployment of a malicious co-located VPS in the cloud where the victim VPS resides. As public clouds are not common in ICSs, the clouds in ICSs can be either private or hybrid. The access needed to private or hybrid clouds can be possible in at least three scenarios. 

In the first scenario, the attack can be originated from the cloud provider targeting the VPS of cloud users \cite{rakotondravony2017classifying}. As cloud providers provide software, platform, and infrastructure as service \cite{annapoorani2018analysis}, they have physical access to target clouds where the victim VPS resides. 

In the second scenario, a malicious insider \cite{ylmaz2018cyber,choi2020expansion}, which can be a disgruntled employee, can use his insider knowledge of the system to deploy the malicious co-located VPS. A similar incident is found in the literature where a disgruntled ex-employee of an ICS posted a note in a hacker journal indicating that his insider knowledge of the system could be used to shut down that ICS \cite{stidham2001can}.

The third scenario is interdiction, which has been rumored to be used in the past ~\cite{spiegel,macri,snyder} and has been recently proven to be practically feasible ~\cite{swierczynski2016interdiction}. In this scenario, during interdiction, a competitor can intercept the installation of VPS in clouds while providing service and may deploy the malicious VPS.

\vspace{0.1em}

\textbf{(v) Stealthy attack:} The authorities may not be aware of the co-located malicious VPS and would possibly not detect the source of our attack. In this sense, our attack is stealthy and can alter the normal behavior of PLCs in ICSs while remaining unidentified.

\vspace{0.0em}
\textbf{(vi) Attacker's cost:} 
Most of these specific DLLs are available as open-source, and very few are proprietary. To acquire the open-source DLL files, 
the attacker has a zero cost. To acquire the DLL files of the proprietary cloud protocols, the attacker just needs to buy a basic commercial license that  may cost a minimum of \$100 \cite{commerciallicence}. Moreover, most proprietary cloud protocols have a free evaluation for few days, and the attacker can also use this free evaluation period to access the .bss section of the target control DLL.





\vspace{-0.30em}
\section{\textbf{.bss section of target control DLL}} 
\label{subsec:Duplicating section of DLLs}

To recreate the .bss imposter page, the attacker first needs to find the target control DLL file of cloud protocols (i.e., MQTT, AMQP) that transports the control commands from the VPS to  PLCs.

\vspace{-0.50em}
\subsection{\textbf{Target control DLL file}}
\label{subsubsec:Target control DLL file}

Mostly, the name of the  target control DLL file depends upon the cloud protocol's implementation variants. For example, the name of a popular implementation of MQTT cloud protocol is Mosquitto, and the target control DLL file for this variant to access by the attacker is mosquitto.dll. 
We do an exhaustive search and tabulate five popular variants of MQTT and their target control DLL files in Table \ref{table:target control DLLS}. The same approach is equally applicable to other cloud protocols. The DLL files are located in the parent directory of the installation folder in the cloud. 

\begin{table}[h!]
\vspace{-0.80em}
	\footnotesize
	\centering
		\caption{Target control DLL file of cloud protocol variants}
		\vspace{-1.520em}
		\label{table:target control DLLS}
		\begin{tabular}{|p{0.4cm}|p{3.50cm}|p{2.5cm}|}
			\hline
			\cellcolor [gray]{0.85} \textbf{Sl.} &  \cellcolor [gray]{0.85} \textbf{Cloud protocol variants} &  \cellcolor [gray]{0.85} \textbf{Target control DLL} \\
			\hline
			\hline
    		1 & EMQ X  Broker \cite{EMQXbroker}& erlexec.dll\\
			\hline
			2 & Mosquitto \cite{mosquitto}  & mosquitto.dll\\
			\hline
			3 & MQTT-C  \cite{mqttc} & mqtt\_pal.dll\\
			\hline
			4 & eMQTT5 \cite{eMQTT} & MQTT\_client.dll\\
			\hline
			5 & wolfMQTT   \cite{wolfmqtt} & MqttMessage.dll\\
			\hline
		\end{tabular}
	\vspace{-01.10em}
\end{table}
\vspace{-01.0em}

\subsection{\textbf{Format of target control DLL files}}
\label{subsubsec:Format of DLL files}

In  64-bit Windows, DLL files follow Portable Executable 32+ (PE32+) format. In high level, PE32+ has a number of \textit{headers} and \textit{sections} (Fig. \ref{fig:bayesian_estimation}). The header consists of DOS header, PE header, optional header, section headers, and data directories. These headers have \textit{Image base Address} and relative virtual address (RVA) of every section that tells the dynamic linker how to map every section of the DLL file into physical memory. There are different sections placed after headers in DLL. Among different sections in DLLs, we want to mention four sections, namely .rdata, .data, .text, and .bss sections. The .rdata section contains string literals, the .data section contains global/static initialized variables, the .text section contains machine code of the program, whereas the .bss section contains zero-initialized variables.  It is important to note that all these sections are \textit{page-aligned} \cite{pietrek1994peering}. This means that these sections must begin on a multiple of a page size in both virtual and physical memory. These sections of DLL files are mapped to pages in physical memory after the \textit{base-relocation} \cite{pietrek1994peering}. The base-relocation is randomized, and the \textit{ASLR technique} is used to map these sections to pages in physical memory at \textit{load time} by the operating system.

\subsection{\textbf{Reasons for choosing the .bss section}}
\label{subsubsec:reasons for choosing the .bss section}

The intention of the attacker is to find a section in the DLL file that has less entropy, which leads to a successful guess of the section. As the .rdata, the .data, and the .text sections consist of different unknown data and addresses, the pages in physical memory corresponding to these three sections have higher entropy. Hence, the estimation of these pages by the attacker requires large memory and time \cite{barresi2015cain} that is not computationally feasible. 

On the other hand, we examine that the .bss section of a target control DLL file of cloud protocols (i.e., MQTT, AMQP) is responsible for transporting control programming and supervisory control-related data, which are static except a new control command is issued. The .bss section contains different uninitialized global/static variables. They are also known as \textit{tag values} and are organized in a tag table. The tag table is typically placed in the .bss section.

\begin{figure}[ht!]
\vspace{-1.210em}
\centering
\includegraphics[width=0.4\textwidth,height=0.14\textheight]{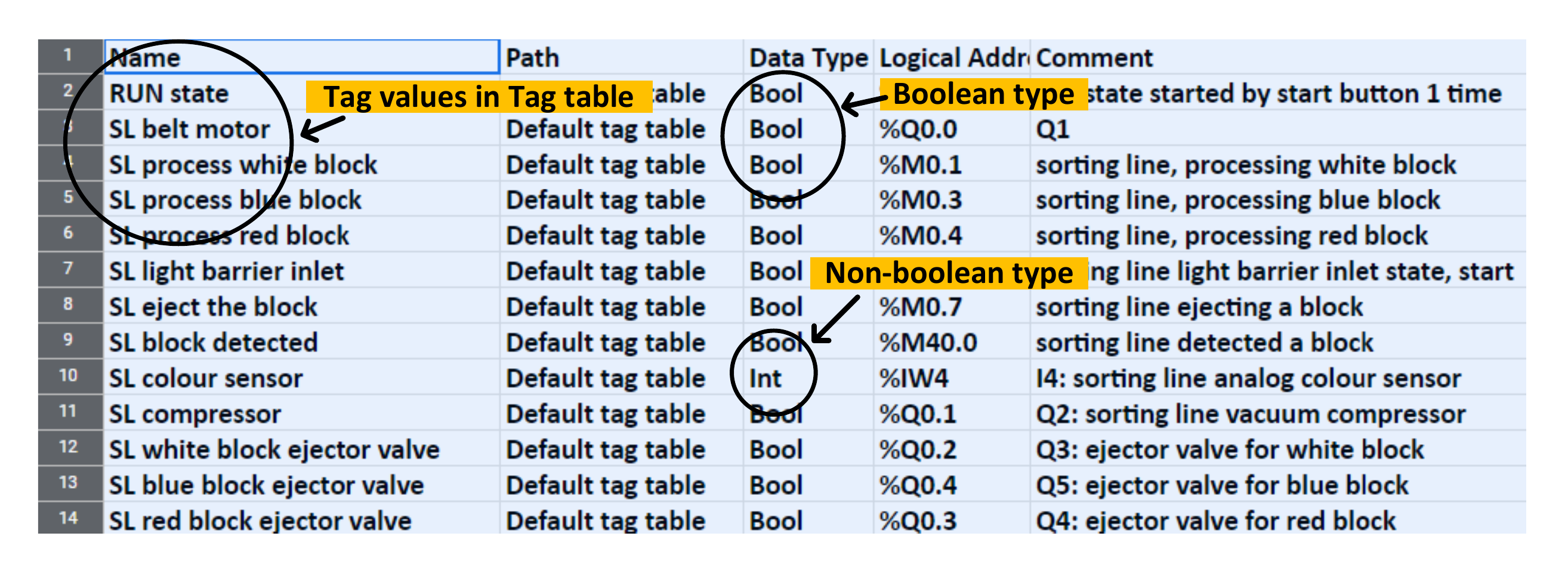}
\vspace{-01.81em}
\caption{Tag values in  tag table of the TIA portal.}
\label{fig:tagvalues_tagtable}
\vspace{-1.0em}
\end{figure}

\textbf{An example of the tag values:} We use a real-world testbed of  an automated high-bay warehouse from \textit{fischertechnik}. The warehouse is connected with a SIMATIC S7-1500 PLC from Siemens. The PLC communicates with the cloud using a TIA portal \cite{TIAportal} through the MQTT cloud protocol Mosquitto.  
A snippet of tag values in the tag table sent from the TIA portal  to the SIMATIC PLC are shown in Fig. \ref{fig:tagvalues_tagtable}. A complete list of the tag values is provided in the following link: {\color{blue}\url{https://sites.google.com/view/bayesmem/home}}.


If we analyze the tag values in tag tables (Fig. \ref{fig:tagvalues_tagtable}), we can observe that tag values correspond to particular states of the target ICS, e.g., the position of a vacuum gripper robot in the warehouse. Most of the tag values are boolean, and very few of them are other data types. The initialization of tag values to either \textit{0 or 1} or non-boolean values in .bss section depends on states of the target ICS and  increases entropy. Therefore, it provides a challenge to the attacker to successfully recreate the .bss section. Thankfully, this challenge can be handled by using the Bayesian estimation of  specific command data in the .bss section. This process is  discussed in the next section.


\section{Bayesian estimation of .bss section}
\label{sec:Bayesian estimation of the .bss section}

We first mathematically model ICSs using the Bayesian estimation and  then use the model to recreate the .bss imposter page. 



\textbf{\textit{Proposition 1- State-space model of an ICS:}} An ICS is dynamic in nature and can be expressed as a discrete-time state-space model \cite{friedland2012control}.  Therefore, a control system in ICS can be expressed by a state vector $x_{k}$, which is a parameter of interest, and a measurement vector $y_{k}$, which is the measurement for  $x_{k}$ at discrete-time index $k$ (see Fig. \ref{fig:bayesian_estimation}). The terms $x_{k}$ and $y_{k}$ can be expressed as:


 \vspace{-0.90em}
\begin{equation}
\begin{aligned}
x_{k} = f_{k-1}(x_{k-1},q_{k-1}) = p(x_{k} | x_{k-1})  
\label{eqn:motionmodelx}
\end{aligned}
\end{equation}
\vspace{-1.50em}

\vspace{-1.0em}
\begin{equation}
\begin{aligned}
y_{k} = h_{k}(x_{k},r_{k}) = p(y_{k} | x_{k})  
\label{eqn:measurementmodely}
\end{aligned}
\end{equation}
\vspace{-1.3em}

where $q_{k-1}$ and $r_{k}$ are state noise and measurement noise vector respectively, and they are mutually exclusive. Please note that both $x_{k}$ and $y_{k}$ are stochastic processes, and Eqn. \ref{eqn:motionmodelx} implies that current state $x_{k}$ at time index $k$ depends on only the previous state $x_{k-1}$ at time index $k-1$ (i.e., Markov process). We implement the state space model of ICS in lines 2-3 of our \textit{BayesImposter} algorithm \ref{alg:bayesmemAlgorithm}.

\textbf{\textit{Source of the data to create the state-space model:}}
To create the state-space model and to estimate $x_k$ and $y_k$, the main challenge for the attacker is to gather the previous states, $x_{1:k-1}$ and previous measurements, $y_{1:k-1}$.  \textit{The attacker can gather  $x_{1:k-1}$ and $y_{1:k-1}$ from OPC tags, historian data, specific PLC block information, or network traffic \cite{choi2020expansion}. Moreover, as mentioned in Section \ref{sec:attack model}, the cloud provider, or a malicious insider, or an interdiction method can make it possible to get $x_{1:k-1}$ and $y_{1:k-1}$ from these sources.} The attacker can use $x_{1:k-1}$ and $y_{1:k-1}$ to create a probabilistic graphical model - Bayes net, which is a directed acyclic graph describing how a joint density can be factorized. The Bayes net also illustrates conditional dependencies among all the states in the ICS (Fig. \ref{fig:bayesian_estimation}).

The tag values located in the .bss section are directly related to the current states ($x_k$) and measurements ($y_k$). Therefore, \textit{BayesImposter} has the following two parts:

\textbf{Part 1.} Estimation of the current states ($x_k$) and measurements ($y_k$) of the state-space model.

\textbf{Part 2.} Estimation of tag values from the estimated $x_k$ and $y_k$.

\vspace{-0.3em}
\subsection{\textbf{Estimation of  states and measurements}}
\label{subsec:Estimation of $x_k$ and $y_k$}

At first, we define the univariate and multivariate ICS to provide background on the design space of the state-space model of ICSs.

\textit{\textbf{Definition 1 (Univariate ICS).}} We define an univariate ICS as where each state $x_k$ has only a single measurement quantity $y_k$ at any time step $k$. 

\textit{\textbf{Definition 2 (Multivariate ICS).}} We define a multivariate ICS as where each state $x_k$ has multiple (i.e., n number) measurement quantities, $[y^1_k, y^2_k,......, y^n_k]$ at any time step $k$.

Practically speaking, an ICS is a mixture of univariate and multivariate state-space models. Therefore, the main challenge for the attacker is to satisfactorily estimate the current state $x_k$ and measurement $y_k$ for both univariate and multivariate ICSs. To handle this challenge, we bring Propositions 2 and 3 to estimate $x_k$ and $y_k$ for a univariate ICS and Propositions 4 and 5 for a multivariate ICS.

\begin{figure}[ht!]
\vspace{-1.320em}
\centering
\includegraphics[width=0.49\textwidth,height=0.19\textheight]{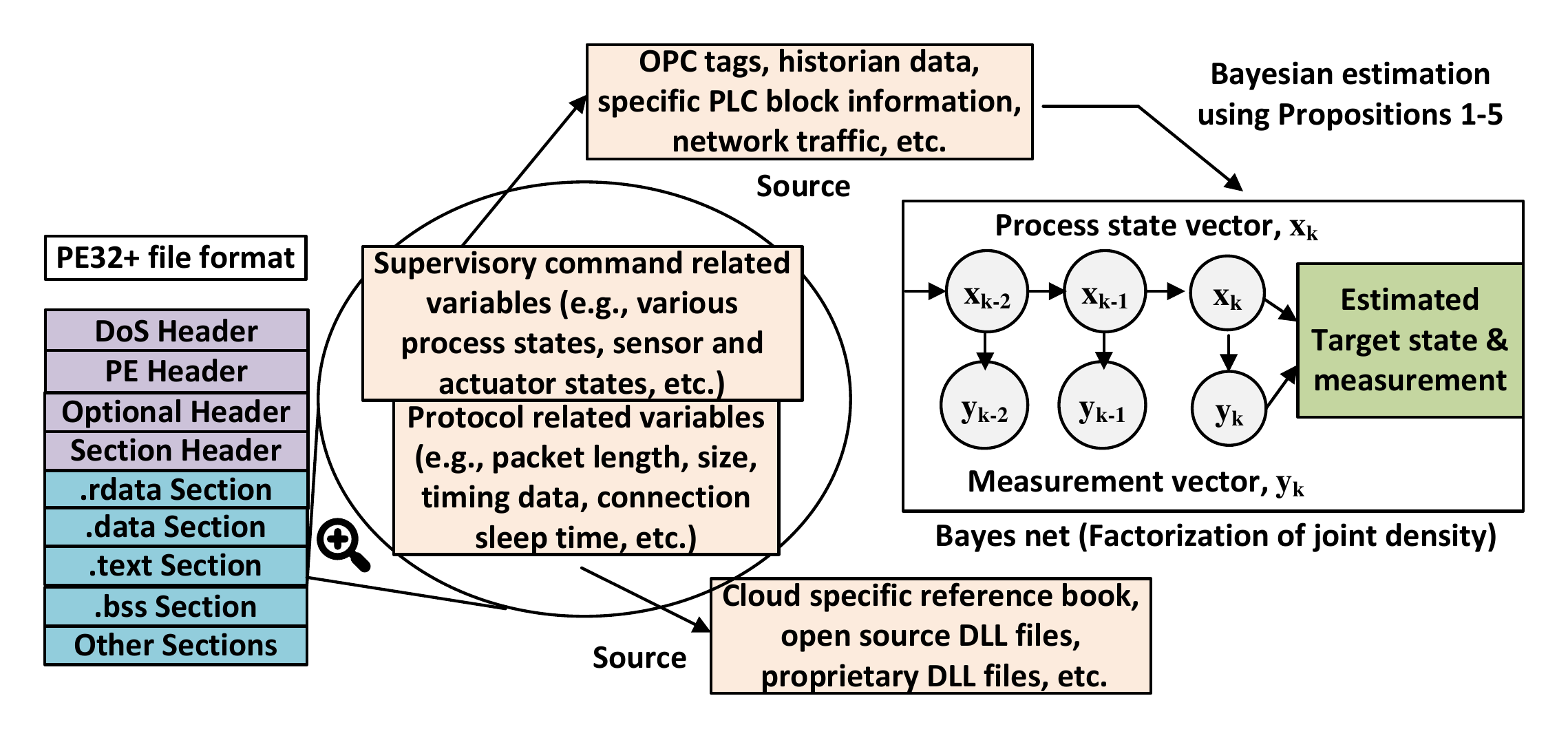}
\vspace{-2.9em}
\caption{An overview of duplicating the .bss section of the target control DLL file.}
\label{fig:bayesian_estimation}
\vspace{-1.10em}
\end{figure}




\textbf{\textit{Proposition 2:}}
\textit{BayesImposter} can predict the current state $x_k$ at time $k$ if the attacker has information only on the previous state $x_{k-1}$ and previous measurements $y_{1:k-1}$, by using the Chapman-Kolmogorov equation. Here, $y_{1:k-1}$ consist of  all previous measurement data $[y_1\, y_2\, ...\, y_{k-1}]$ up-to time $k-1$.

\textbf{\textit{Explanation of Proposition 2:}}
Let us give an example to clear this concept. Let us denote the \textit{states} of a suction cup of the vacuum gripper robot in our example warehouse as $x_k$ at time $k$. Let us consider the suction cup can be in one of two states, $x_k \epsilon \{\text {ON, OFF\}}$. The activation of the suction cup in each state depends on the position of the horizontal and vertical axis of the vacuum gripper robot (see Fig. \ref{fig:synergy_ics}). The position measurement can be expressed by $y_k$ at time $k$. If the attacker knows previous state $x_{k-1}$ of the suction cup and previous position measurements  $y_{1:k-1}$, then the attacker can use these data to accurately estimate the current state $x_k$ at time $k$ by using Eqn. \ref{eqn:chapmankolmogorov} (i.e., Chapman-Kolmogorov equation). The L.H.S of Eqn. \ref{eqn:chapmankolmogorov}, $p(x_{k} | y_{1:k-1})$,  is a conditional estimation of current state $x_k$, while previous measurements $y_{1:k-1}$ are given. The R.H.S of Eqn. \ref{eqn:chapmankolmogorov} depicts that  $p(x_{k} | y_{1:k-1})$  is a function of previous state, $x_{k-1}$, that is an indication of Markov process. The Proposition 2 is implemented in lines 6-7 of our \textit{BayesImposter} algorithm \ref{alg:bayesmemAlgorithm}.

\vspace{-01.30em}
\begin{equation}
\begin{aligned}
p(x_{k} | y_{1:k-1}) 
&= \int p(x_{k} | x_{k-1}) p(x_{k-1} | y_{1:k-1}) dx_{k-1}
\label{eqn:chapmankolmogorov}
\end{aligned}
\end{equation}
\vspace{-01.0em}

\textbf{An example:} The name of a specific tag value in the .bss section of the mosquitto.dll is \texttt{suctionstate}, which corresponds to the state information $x_k\epsilon \{\text {ON, OFF\}}$ of the suction cup of our example automated high-bay warehouse. After estimating the state $x_k$ using Eqn. \ref{eqn:chapmankolmogorov}, the  attacker can initialize the tag value to 0 or 1 of the variable \texttt{suctionstate} in the .bss section. If the .bss section contains multiple uninitialized tag values originating in the VPS, the attacker can use a similar technique to successfully estimate all uninitialized tag values and can recreate the .bss section.


\textbf{\textit{Proposition 3:}}
\textit{BayesImposter} can predict the current measurement $y_k$ if the attacker has information on current state $x_{k}$. 

\textbf{\textit{Explanation of Proposition 3:}}
It  is  important  to  note  that  along with state  information $x_k$,  the .bss section transports current measurement $y_k$ from VPSs to PLCs. The importance of sending  measurement information $y_k$ from VPSs to PLCs is explained below. 

\textbf{An example:} In the automated high-bay warehouse, a solenoid is present in the suction cup of the vacuum gripper robot that is turned on/off if the position of the horizontal and vertical axis is above/below a threshold position. Let us denote this threshold position by $S_{\theta}$. If the threshold position is required to be  changed by the upper management located in the cloud, the VPS can send a new threshold position $S^{\theta}_k$ to overwrite the previous value $S^{\theta}_{k-1}$. The new threshold position $S^{\theta}_k$ is equivalent to the current measurement $y_k$, which depends on the current state $x_k$ of the suction cup. Therefore, the current measurement, $y_k = S^{\theta}_k$, can be calculated using the Naive Bayes estimation equation as below:

\vspace{-01.00em}
\begin{equation}
\begin{aligned}
p(y_{k} = S^{\theta}_k | x_{k}) 
&= \dfrac{p(x_{k}|y_{k} = S^{\theta}_k) \times p(y_{k} = S^{\theta}_k) } {\sum_{y_k} p(y_k)p(x_k | y_k)}
\label{eqn:bayesequation}
\end{aligned}
\end{equation}
\vspace{-0.10em}

Here, the likelihood term, $p(x_{k}|y_{k} = S^{\theta}_k)$, is calculated from the frequency distribution of the measurement $y_k$ for the state $x_k$. The frequency distribution is calculated from the OPC tags and the historian data (Fig. \ref{fig:bayesian_estimation}). The prior probability, $p(y_{k} = S^{\theta}_k)$, is the probability that the parameter takes on a particular value $S^{\theta}_k$, prior to taking into account any new information (i.e., current state $x_k$). If the probability of the estimation, $p(y_{k} = S^{\theta}_k | x_{k})$, is below a cut-off value ($K_c$),  \textit{BayesImposter} discards that estimation and picks another $y_k = S^\theta_k$ to test in Eqn. \ref{eqn:bayesequation}. By this way, the attacker can use \textit{BayesImposter} to estimate any measurement quantity $y_k$ at time step $k$. It is noteworthy that if the current state $x_k$ is unknown, \textit{BayesImposter} can use the Proposition 2 to calculate the current state $x_k$ first, and then use the Proposition 3 to calculate $p(y_k|x_k)$ using Eqn. \ref{eqn:bayesequation}. The Proposition 3 is implemented in lines 9-17 of our proposed \textit{BayesImposter} algorithm \ref{alg:bayesmemAlgorithm}.

\textbf{\textit{Proposition 4:}}
If multiple (i.e., n) measurement quantities, [$y^1_k$, $y^2_k$, $y^3_k$,..., $y^n_k$], at a time step $k$, jointly contribute to estimate any state $x_k$, \textit{BayesImposter} uses the joint probability of multiple measurement quantities, $p(y^1_k \cap y^2_k \cap y^3_k \cap......\cap y^n_k)$, in Eqn. \ref{eqn:chapmankolmogorov}. 

\textbf{\textit{Explanation of Proposition 4:}}
Let us assume that each state $x_k$ in a multivariate ICS has n number of measurements at every time step. For example, at state $x_1$, the ICS has $y^1_1, y^2_1, y^3_1,......, y^n_1$ measurement values; at state $x_2$, the ICS has $y^1_2, y^2_2, y^3_2,......, y^n_2$ measurement values and so forth. Let us denote the joint probability of n number of measurement values at state $x_k$ by $Y_k = p(y^1_k \cap y^2_k \cap y^3_k \cap......\cap y^n_k)$. Eqn. \ref{eqn:chapmankolmogorov} is modified in the following way to accommodate the joint probability of measurement values.

\vspace{-1.20em}
\begin{equation}
\begin{aligned}
p(x_{k} | Y_{1:k-1}) 
&= \int p(x_{k} | x_{k-1}) p(x_{k-1} | Y_{1:k-1}) dx_{k-1}
\label{eqn:chapmankolmogorovjointprobabilty}
\end{aligned}
\end{equation}
\vspace{-1.200em}

where joint probability of measurement values from time step 1 to $k-1$ is denoted by $Y_{1:k-1}$. The Proposition 4 is implemented in lines 20-22 of our proposed \textit{BayesImposter} algorithm \ref{alg:bayesmemAlgorithm}. 

\textbf{An example:} From the explanation of the Proposition 2, we know that the suction cup can have any one of the following two states: $\{ON,OFF\}$, depending upon the position of the horizontal and vertical axis of the vacuum gripper robot. In multivariate ICS, instead of having a single position value for a particular state, the horizontal and vertical axis could have multiple position values within a range. For example, a position within 0 cm to 10 cm  of the horizontal axis could trigger the  state to ON from OFF. If there are n measurement values within the position range of 0 cm to 10 cm, \textit{BayesImposter} uses Eqn. \ref{eqn:chapmankolmogorovjointprobabilty} to estimate the next state $x_k$.

\setlength{\textfloatsep}{0pt}
\vspace{-0.00em}
\begin{algorithm}[ht!] 
	\footnotesize
	\DontPrintSemicolon
	\caption{BayesImposter Algorithm.}
	\label{alg:bayesmemAlgorithm}
	\KwIn{Previous measurements, $y_{1:k-1}$ and states $x_{1:k-1}$ up to k-1
	}
	\KwOut{Current measurements, $y_k$ and states, $x_k$ at k step
	} 
	\For (\tcp*[h]{Proposition 1  for state-space model}){$k \gets 1$ \textbf{to} k-1} { 
		Collect $y_{1:k-1}$ and $x_{1:k-1}$ information of ICS\\
		Create state-space model: $x_k = p(x_k|x_{k-1})$ \& $y_k = p(y_k|x_k)$\\
		\If{ICS is univariate}{
		     \For(\tcp*[h]{Proposition 2  for $x_k$}){Each unknown $x_k$} { 
		          Find $p(x_k|y_{1:k-1})$ for every $x_k$ \\
		          Select $x_k$ having the highest  $p(x_k|y_{1:k-1})$\\
		     }
		     \For(\tcp*[h]{Proposition 3  for $y_k$}){Each unknown $y_k$} { 
		                 \If{$x_k$ is known}{
		                         Find $p(y_k|x_k)$ for every $x_k$\\
		                         \If{$p(y_k|x_k)$ $>$ cut-off $K_c$}{
		                             Select the $y_k$ as the estimation\\
		                         }
		                         \Else{
		                             Discard the estimated $y_k$\\
		                         }
		                 }
		                 \Else{
		                     Find $x_k$ first using Proposition 2\\
		                     Then use Proposition 3\\
		                 }
		     }
		}
       \If{ICS is multivariate}{
          \For(\tcp*[h]{Proposition 4  for $x_k$}) {Each unknown $x_k$} { 
		           Find joint probability $Y_k = p(y^1_k \cap y^2_k \cap......\cap y^n_k)$\\
		          Find $p(x_k|Y_{1:k-1})$ for every $x_k$\\
		          Select $x_k$ having the highest  $p(x_k|Y_{1:k-1})$\\
		     }
		  \For(\tcp*[h]{Proposition 5  for $y_k$}){Each unknown $y_k$} { 
		                 \If(\tcp*[h]{$\max\limits_{\forall y_k}$ function}){$x_k$ is known}{
		                         Find $p(y^1_k|x_k)$ for $y_k\epsilon\{y^1_k,y^2_k,..,y^n_k\}$\\
		                         $max$ $\gets$ $p(y^1_k|x_k)$\\
		                          \For{Every $y_k\epsilon\{y^2_k,y^3_k,..,y^n_k\}$} { 
		                                Find $p(y_k | x_k)$\\
		                                \If{ $p(y_k | x_k)$ $>$ max}{
		                                     $max$ $\gets$ $p(y_k|x_k)$\\
		                                }
		                          }
		                          Select $max$ as the $y_k$ for given $x_k$
		                 }
		                 \Else{
		                     Find $x_k$ first using Proposition 2\\
		                     Then use Proposition 5\\
		                 }
		     }
       }
	}
	\vspace{-0.7540em}
\end{algorithm}
\vspace{.00em}

\textbf{\textit{Proposition 5:}}
If multiple (i.e., n) measurement quantities, [$y^1_k$, $y^2_k$, $y^3_k$, ......, $y^n_k$], at a time step $k$, present in a multivariate ICS, \textit{BayesImposter} finds $y_k$ that gives the highest probability in Eqn. \ref{eqn:bayesequation}. 

\textbf{\textit{Explanation of Proposition 5:}}
The Proposition 5 is an extension of the Proposition 3 for multiple number of measurement values $[y^1_k, y^2_k, y^3_k,......, y^n_k]$, at a current state $x_k$. To estimate a measurement value from multiple measurement values, \textit{BayesImposter} plugs in most frequent values from the distribution of measurement values $[y^1_k, y^2_k, y^3_k,......, y^n_k]$ in Eqn. \ref{eqn:bayesequation} with an intention to maximize the left hand side of Eqn. \ref{eqn:bayesequation}. For example, if the threshold position in the explanation of Proposition 3 has multiple values $S^{\theta1}_k$, $S^{\theta2}_k$,...,$S^{\theta n}_k$  for current state $x_k$, we can write Eqn. \ref{eqn:bayesequation} as below.

\vspace{-01.100em}
\begin{equation}
\begin{aligned}
\max_{\forall y_k} \,\,\{p(y_{k} | x_{k})\}
&= \max_{\forall y_k} \,\,\{ \dfrac{p(x_{k}|y_{k}) \times p(y_{k}) } {\sum_{y_k} p(y_k)p(x_k | y_k)}\}
\label{eqn:bayesequationmultivariate}
\end{aligned}
\end{equation}
\vspace{-01.10em}

where $y_k \epsilon \{ S^{\theta1}_k, S^{\theta2}_k,...,S^{\theta n}_k \}$. The $\max\limits_{\forall y_k}$ is the function that maximizes $p(y_{k} | x_{k})$ for all $y_k$ that is implemented using an iterative approach in lines 24-34 of the proposed \textit{BayesImposter} algorithm \ref{alg:bayesmemAlgorithm}.


\vspace{-0.2em}
\subsection{\textbf{Tag values from the estimated $x_k$ and $y_k$}}

It is mentioned earlier in section \ref{subsec:Duplicating section of DLLs} that the .bss section contains different uninitialized global/static tag variables. They can be broadly divided into two categories, namely the control programming or command related variables and protocol related variables (Fig. \ref{fig:bayesian_estimation}). 

\textit{\textbf{Estimation of control commands from $x_k$ and $y_k$:}}
After estimating $x_k$ and $y_k$, the next challenge is to look for the corresponding control commands from the estimated $x_k$ and $y_k$.  It can be done in two ways. \textit{Firstly}, most  control commands are the direct values of $x_k$ and $y_k$ that are already estimated by \textit{BayesImposter}. For example,  from the Proposition 2, the threshold position $S^{\theta}_k$ is equal to the estimated measurement $y_k$ in the .bss section.  \textit{Secondly}, rest of the control commands are estimated from OPC tags and specific PLC information (Fig. \ref{fig:bayesian_estimation}) using the estimated $x_k$ and $y_k$. For example, the value of \texttt{suctionstate} $\epsilon\{ON,OFF\}$ corresponding to 0 or 1 can be found from specific PLC information (see Section \ref{subsec:Entropy in the .bss section}).


\textit{\textbf{Estimation of protocol related variables:}}
The protocol-related variables are specific to cloud protocols and hence, are fixed and initialized at the load time of the control DLL file. 
The attacker can get the list of all the protocol-related variable names and their values from the reference book of a specific cloud protocol. As mentioned in Section \ref{sec:attack model}, most of the target control DLLs  are available as open-source, and very few are proprietary, which are accessible by a basic commercial license (cost less than \$100 \cite{commerciallicence}).

\vspace{-0.6em}
\subsection{\textbf{Entropy in the .bss section}}
\label{subsec:Entropy in the .bss section}

The size of the specific control variable used in the .bss section can be a maximum of 64 bits in a 64-bit machine. Therefore, we have an entropy of $2^{64}$ possible values. For example, the tag variable \texttt{suctionstate} ideally could have $2^{64}$ values. But, in real-world implementation, the control variables are problem-specific and they have very few key values, which are also problem specific. Therefore, as mentioned in Proposition 2,
the state variable, \texttt{suctionstate}, has two possible key values: $\{\text {ON, OFF}\}$. So, the entropy of the \texttt{suctionstate} is not $2^{64}$; instead, the entropy is only two. Moreover, these key values are declared in the header files of the program codes, and programmers, as a good practice, generally use user-defined data types, such as \textit{Enumeration (enum)} type to declare these key values. The use of enum data type by the programmer makes the declared control variable (e.g., \texttt{suctionstate}, etc.) more predictable. For example, after careful examination of control-related application codes that are running on top of cloud protocols, we find the following code snippet that supports our observation:  

\vspace{0.0em}
{
\footnotesize
\noindent $\texttt {enum statepool \{0,1\};}$\\
\noindent $\texttt {enum statepool suctionstate;}$
}
\vspace{0.0em}

This indicates that the values of ON/OFF is 0 or 1. In this way, the attacker can specifically know the tag values in the .bss section to recreate the .bss imposter page. 

\vspace{-0.21em}
\section{Memory Deduplication+Rowhammer}
\label{subsec:Merging the duplicated .bss section}


So far, we have discussed how the attacker can recreate the .bss imposter page using \textit{BayesImposter}. Now, we discuss how the attacker uses the memory deduplication + Rowhammer bug to trigger a bit flip in the recreated .bss imposter page to corrupt control commands. 
As recent works \cite{razavi2016flip, barresi2015cain, bosman2016dedup,oliverio2017secure} have already provided details on the memory deduplication + Rowhammer bug, we will not repeat the same details here. We refer to Appendix \ref{appendix:sec:Memory deduplication + Rowhammer} for more details. 
Instead, we provide advantages of our approach over \cite{razavi2016flip, barresi2015cain, bosman2016dedup,oliverio2017secure}. Let us briefly discuss the memory deduplication + Rowhammer first.

\begin{figure}[ht!]
\vspace{-1.320em}
\centering
\includegraphics[width=0.48\textwidth,height=0.15\textheight]{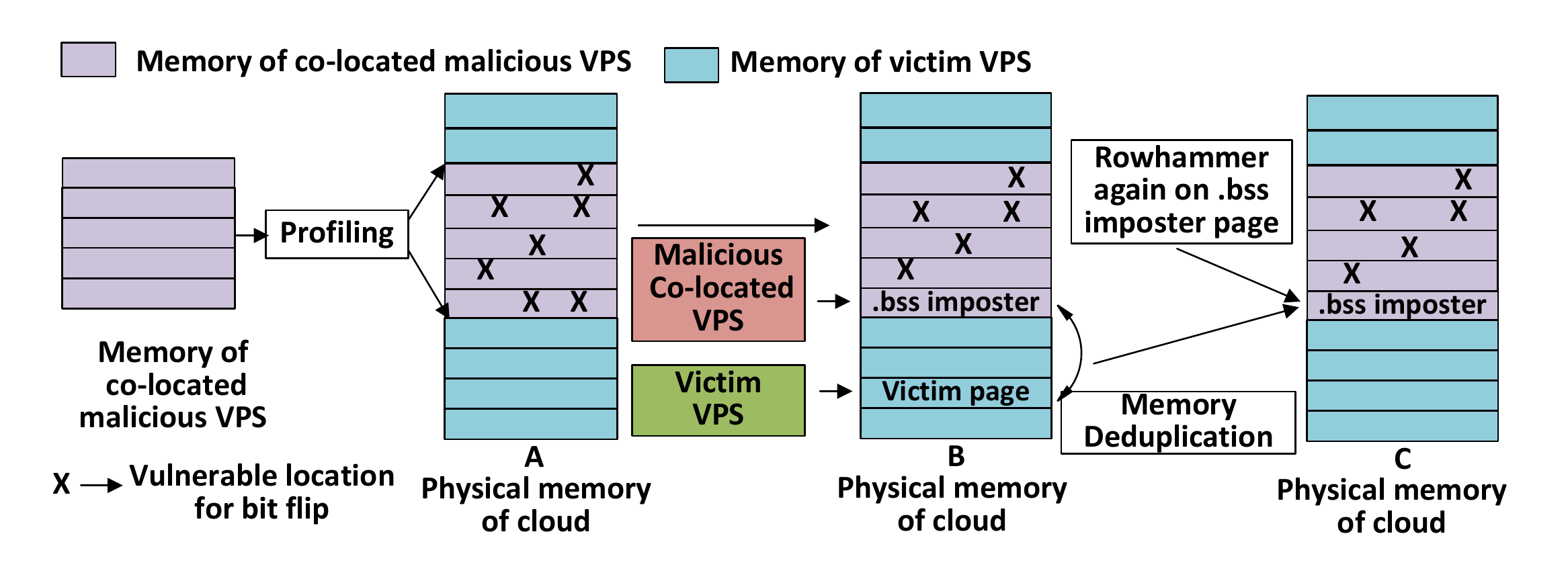}
\vspace{-2.4905em}
\caption{(A) Profiling the memory of cloud. (B) Placing \textit{.bss imposter page} in the vulnerable location. (C) After memory deduplication, \textit{victim page} is backed by the \textit{.bss imposter page} and the Rowhammer causes bit flips in the  \textit{.bss imposter page}.}
\label{fig:memorydedup_rowhammer}
\vspace{-1.00em}
\end{figure}

\textit{\textbf{Brief overview:}}
Memory deduplication  merges identical pages located in the physical memory into one page. Rowhammer \cite{kim2014flipping} is a widespread vulnerability in recent DRAM devices in which repeatedly accessing a row can cause bit flips in adjacent rows. 

Memory deduplication thread (i.e., KSM) running in the host cloud hypervisor (i.e., KVM in Linux) maintains stable/unstable trees in a red-black tree format to keep track of the pages having identical contents in memory. If the \textit{.bss imposter page} arrives first in the memory provided from the co-located malicious VPS, the node of the red-black tree will be updated first with the \textit{.bss imposter page}. Therefore, if the \textit{victim page} comes later from the victim VPS, the \textit{victim page} is merged with the \textit{.bss imposter page}, and the \textit{victim page} shares the same memory location of the \textit{.bss imposter page}. In this way, the attacker can control the memory location of the \textit{victim page} and can trigger a Rowhammer on that page.

The first step to initiate Rowhammer is to find the \textit{aggressor/victim} addresses in the physical memory of the running system. This step is named as \textbf{profiling}. The \textit{aggressor} addresses are the memory locations within the process's virtual address space that are hammered, and the \textit{victim} addresses are the locations where the bit flips occur (Fig. \ref{fig:memorydedup_rowhammer}(A)). From the profiling step, the attacker knows the aggressor rows for the vulnerable memory locations. After placing the \textit{.bss imposter page} in one of the vulnerable locations, the attacker hammers again on the aggressor rows (Fig. \ref{fig:memorydedup_rowhammer}(C)). This results in bit-flips in the \textit{.bss imposter page} that in effect changes the control commands in the .bss section of the target control DLL.

\vspace{-0.750em}
\subsection{\textbf{Advantages of BayesImposter}} 
\label{Difference between BayesMem and recent works}

\subsubsection{\textbf{No first precedence and two copies of target pages}}
\label{subsubsec:BayesMem is more practical, stronger, and reliable}

To ensure that the \textit{.bss imposter page} arrives first in the memory, the attacker's VPS should start first before the victim VPS. This is known as the first precedence. Recent works \cite{razavi2016flip, barresi2015cain, bosman2016dedup,oliverio2017secure} use this technique along with creating two copies of target pages to place the \textit{.bss imposter page} in the red-black tree before the target \textit{victim page}. These techniques require more control over the victim VPS and may not be feasible in practical ICSs. For example, the attacker may not know when the victim VPS is started.

Thanks to the Bayesian estimation of the  \textit{victim page}. Referring to Section \ref{sec:Bayesian estimation of the .bss section}, if the attacker can predict the current states ($x_k$) and measurements ($y_k$), this means that he actually can predict the \textit{victim page} before time $k$. 
As the attacker has the predicted \textit{victim page}, the attacker can provide this predicted \textit{victim page} to the memory deduplication thread at any time. Hence, the attacker does not need to start his VPS before the victim or does not need to create two copies of the target pages in our attack model. This makes our attack model more practical and reliable in the context of ICSs. 

\vspace{-0.61em}
\subsubsection{\textbf{BayesImposter provides simpler profiling step}}
\label{subsubsec:BayesMem provides more refined profiling step}

Recent works \cite{razavi2016flip, barresi2015cain, bosman2016dedup,oliverio2017secure} activate the \textit{large pages} \cite{largepage} in VPS to exploit the double-sided Rowhammering. However, \textit{large pages}  
may not be explicitly turned on in the victim VPS. Therefore, double-sided Rowhammering may not be feasible in the context of ICSs \cite{seaborn2015exploiting}. Therefore, \textit{BayesImposter} uses the random address selection approach for profiling the bit-flippable memory locations. 

In this approach, \textit{BayesImposter} allocated a 1 GB block of memory using a large array filled with doubles. A value of $1.79769313486231 \times 10^{308}$ is stored as double that gives 1 in  memory locations. Next, the attacker randomly picks virtual aggressor addresses from each page of this large memory block and reads $2\times10^6$ times. Then  the attacker moves to the next page and repeats the same steps. As the attacker can know the number of memory banks of the running system from his VPS, he can calculate his chance of hammering addresses in the same bank. For example, in our experimental setup, the machine has 2 Dual Inline Memory Modules (DIMMs) and 8 banks per DIMM. Therefore, the machine has 16 banks, and the attacker has a 1/16 chance to hit aggressor rows in the same bank. 
Moreover, the attacker hammers 4 aggressor rows in the same iteration that increases the chance of having successful Rowhammering.


\vspace{-0.210em}
\section{\textbf{Attack model evaluation}}
\label{subsec:Performance analysis of BayesMem}

\subsection{Automated high-bay warehouse testbed}
\label{sec:Modelling a typical cooling system}

We prepare a testbed to evaluate \textit{BayesImposter} on a practical ICS. We choose a scaled-down model of an automated high-bay warehouse (AHBW) from \textit{fischertechnik} connected with a vacuum gripper robot (VGR), multiprocessing oven (MPO), and sorting line (SL). The process begins first in MPO with a workpiece placed in the oven feeder. The processed workpiece from the MPO is then sent to SL using a conveyor belt. The SL sorts the workpiece depending upon color and stores it in the storage location. Next, the VGR uses its suction cup to hold the workpiece and transports it from the storage location to the pre-loading zone of the rack feeder of the AHBW. Then the rack feeder stores the workpiece in the warehouse. A video demonstration of the factory system is given here: {\color{blue}\url{https://sites.google.com/view/bayesmem/home}}.

\begin{figure}[ht!]
\vspace{-1.6310em}
\centering
\includegraphics[width=0.47\textwidth,height=0.22\textheight]{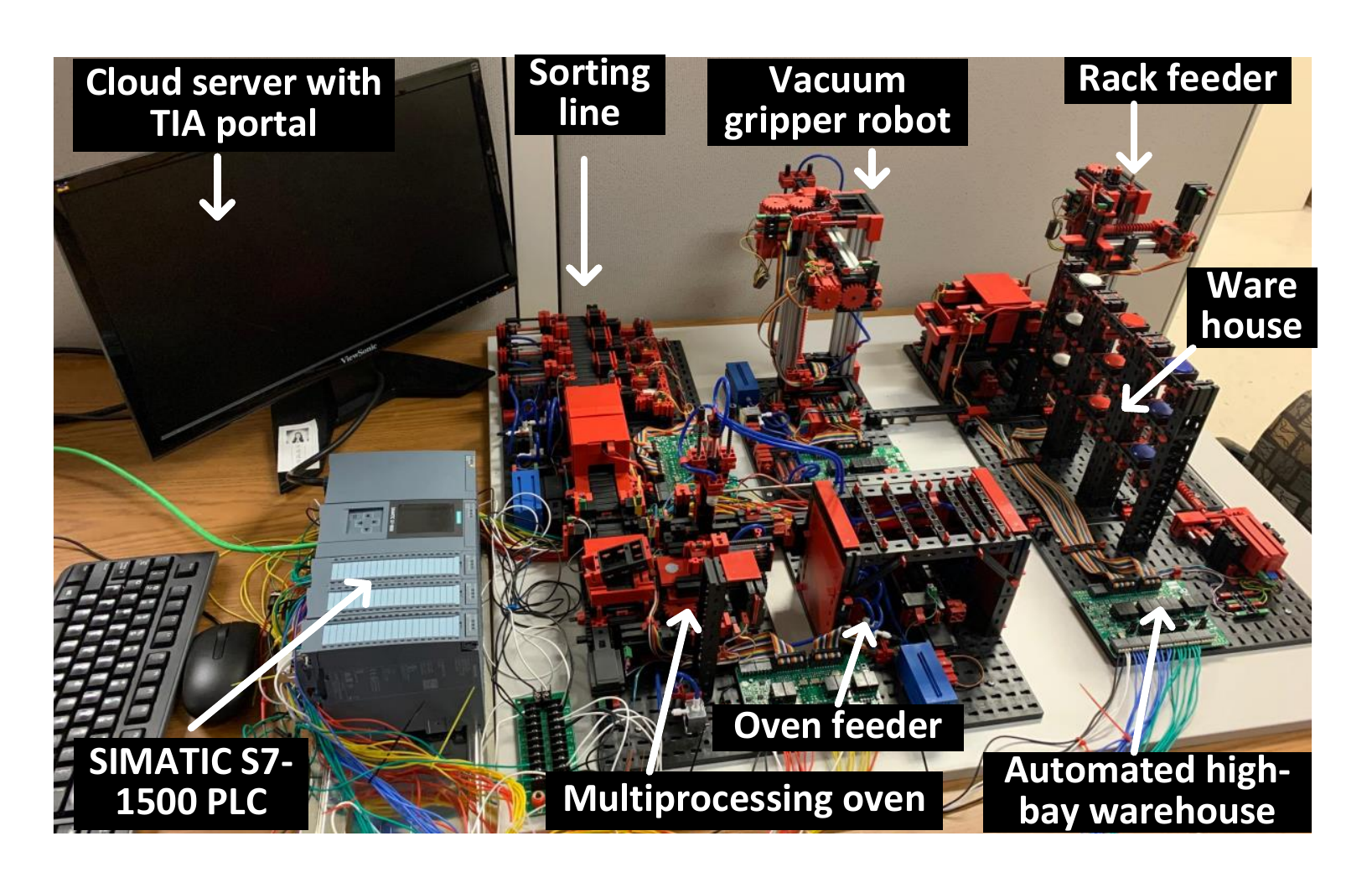}
\vspace{-1.90em}
\caption{A small scale real-world testbed of automated high-bay warehouse to evaluate \textit{BayesImposter}.}
\label{fig:HIL}
\vspace{-01.200em}
\end{figure}

The AHBW is connected with a SIMATIC S7-1500 PLC from Siemens using 32 input/output ports and 8 analog input ports. The PLC communicates with the cloud using a TIA portal through the MQTT cloud protocol Mosquitto. The cloud server runs on Intel CPU i7-6900K with 8 cores and 64GB of DDR3 RAM. We use Ubuntu Server 14.04.2 LTS x86\_64 as the cloud server, which has a Kernel-based Virtual Machine (KVM). Memory deduplication is implemented as Kernel Samepage Merging (KSM) in KVM.  The KVM is kept at its default configuration. The parameters for KSM (see Appendix \ref{subsubsec:Use of Kernel Samepage Merging}) are also kept at their default settings. \textit{All VPSs run with Windows 10} \cite{kSM1000linux} and have 2 GB of main memory. \textit{The idea of \textit{BayesImposter} is equally applicable to the Linux VPSs with .so file \cite{barresi2015cain} of cloud protocols.} The victim VPS is using MQTT to communicate with the PLC using TIA portal. The testbed is shown in Fig. \ref{fig:HIL}.




\vspace{-0.75em}
\subsection{\textbf{Estimation accuracy of \textit{BayesImposter}}}
\label{subsec:Estimation accuracy of the bayesmem}

A practical ICS could have hundreds of states ($x_k$) and measurement values ($y_k$). 
Let us mathematically formulate this first.

\textbf{\textit{Proposition 6:}}
If an ICS has $M$ state variables ($x_k$) and each state variable has $N$ probable states, $N^M$ combinations are possible among state variables and probable states. Similarly, If an ICS has $P$ measurement variables ($y_k$) and each measurement variable has $Q$ probable values, $Q^P$  combinations are possible among measurement variables and probable values. 

After counting, we find that our testbed - automated high-bay warehouse has $M = 420, N = 3, P = 160, Q = 4$. We find that the estimation accuracy for next states or next measurements using Propositions 1-5 of our \textit{BayesImposter} algorithm is  $\sim$91\%. 
It means that \textit{BayesImposter} can estimate the next state or measurement variables within 1/0.91 = 1.09 attempt.

\begin{table}[h!]
\vspace{-1.0em}
	\footnotesize
	\centering
		\caption{Estimation accuracy of \textit{BayesImposter}.}
		\vspace{-01.4250em}
		\label{table:estimationaccuracy}
		\begin{tabular}{|p{3.3cm}|p{4.3cm}|}
			\hline
			\cellcolor [gray]{0.85}Estimating state variables $x_k$ & \cellcolor [gray]{0.85} Estimating measurement variables $y_k$ \\
			\hline
			\hline
			90.2\% & 91.47\%\\
			\hline
		\end{tabular}
	\vspace{-1.0em}
\end{table}
\vspace{-0em}

\vspace{-0.50em}
\subsection{Recreating the .bss imposter page}
\label{sec:Recreating .bss imposter page}

The automated high-bay warehouse testbed has $M = 420$ state variables ($x_k$) in total, and each state has an average of $N = 3$ probable states. The brute-force approach gives $3^{420} \approx 2.4 \times 10^{200}$ combinations according to the Proposition 6. Moreover, this ICS in hand has also $P = 160$ measurement variables ($y_k$) in total, and each variable has an average of $Q = 4$ probable values. The brute-force approach gives $4^{160} \approx 2.13 \times 10^{96}$ combinations. In combined, there are $2.4 \times 10^{200} + 2.13 \times 10^{96} = 2.4 \times 10^{200}$ combinations are possible for the ICS in hand. For a 4KB page size, this may require $(4 \times 2.4 \times 10^{200})$ KB $= 9.6 \times 10^{194}$ GB of guessed pages. In other words, the attacker may need to spray $9.6 \times 10^{194}$ GB pages in the physical memory for successful memory deduplication that is not possible in terms of time and memory. It is not possible to accommodate  $9.6 \times 10^{194}$ GB pages in one attempt of the attack, and the attacker may require thousands of attempts to spray the memory with the guessed pages. In contrast, as \textit{BayesImposter} has an estimation accuracy of $\sim$91\% (see Section \ref{subsec:Estimation accuracy of the bayesmem}), it does not require to guess $N^M$ or $Q^P$ combinations; instead, it can guess states and measurement variables in $1/0.91 = 1.09$ attempt. Therefore, most of the time, \textit{BayesImposter} requires only one or two pages (because of $\sim$91\% accuracy) of size 4KB to spray in the physical memory. 

The victim VPS in our example ICS has a 2 GB main  memory, and it takes $\sim$13  minutes  to  scan  all  the  pages  of  main  memory in a single attempt (see Section \ref{subsec:Attack time}). And, out of 2 GB of memory, we can spray 1.2 GB with the guessed pages at each attempt (i.e., remaining 0.8 GB for operating systems and other applications). Therefore, brute force requires $(9.6 \times 10^{194}) /1.2 = 8 \times 10^{194}$  attempts, whereas \textit{BayesImposter} requires only a 1.09 attempt. As each attempt takes $\sim$13 minutes, \textit{BayesImposter} requires only  $\sim$13 minutes compared to $9.6 \times 10^{194} \times 13$ min.$\,=\, 2 \times 10^{194}$ hours of brute force approach which is not feasible. This reduction of attempts also reduces the attack time (see Section \ref{subsec:Attack time}). As the attack time for \textit{BayesImposter} is significantly low compared to a brute force approach, \textit{BayesImposter} gives more control over the ICS from the attacker's perspective. Table \ref{table:bayesmemeperformance} shows the memory and time requirements for brute-force and \textit{BayesImposter} approaches. 

\vspace{-01.0em}
\begin{table}[h!]
\centering
\caption{Attack time of \textit{BayesImposter}}
\vspace{-01.30em}
\label{table:bayesmemeperformance}
     \begin{tabular}{| p{1.950cm} | p{1.100cm} | p{02.0cm}|p{01.70cm}|} 
     \hline
      \multicolumn {2}{|c|}{\cellcolor [gray]{0.85}BayesImposter} & \multicolumn {2}{c|} {\cellcolor [gray]{0.85} Brute force} \\
     \hline
     \hline
     Guessed page &   Time & Guessed page &   Time \\
     \hline
      4KB or 8KB  & 13 min. & $9.6 \times 10^{194}$ GB & $2 \times 10^{194}$ Hr.\\
     \hline
     \end{tabular}
     \vspace{-01.40em}
\end{table}






 \vspace{-0.20em}
\subsection {\textbf{Attacking the vacuum gripper robot (VGR)}}
\label{subsec:Attacking the solenoidstateBypasshose}
 \vspace{-0.0em}

\begin{figure}[ht!]
 \vspace{-1.50em}
\centering
\includegraphics[width=0.4\textwidth,height=0.12\textheight]{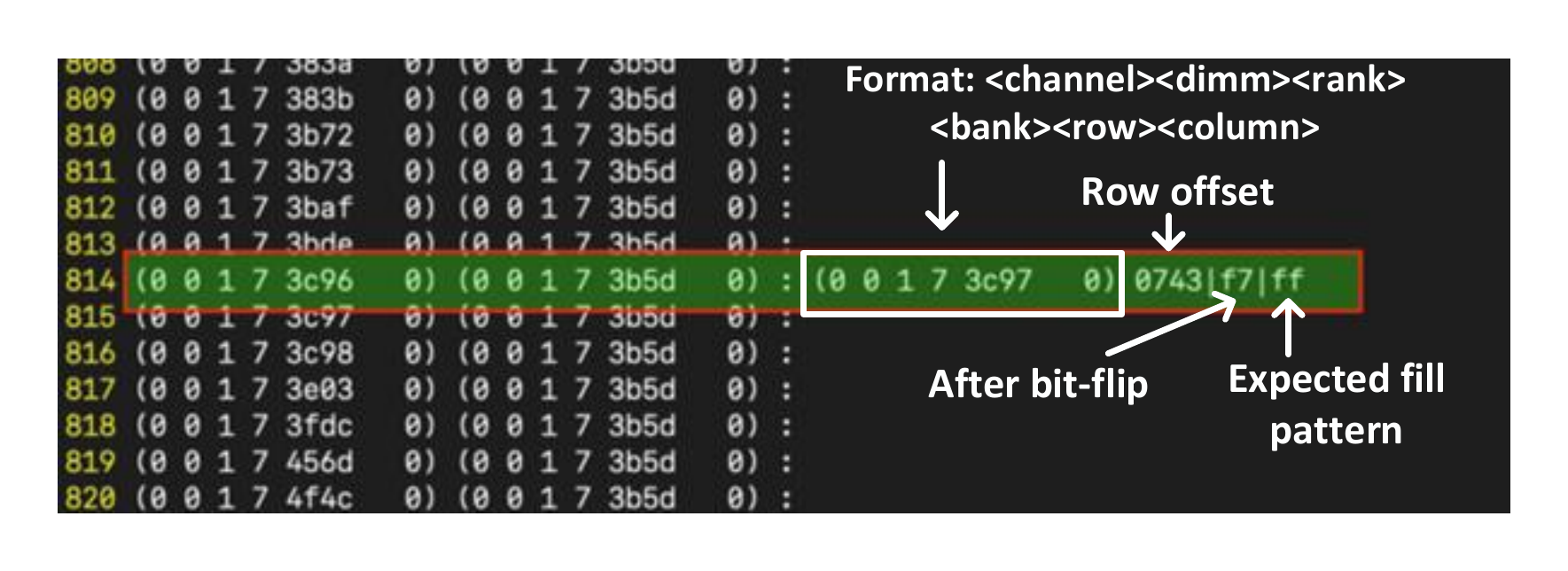}
\vspace{-1.82em}
\caption{Bit-flip in the .bss imposter page.}
\label{fig:rowhammer_bit_flip}
\vspace{-01.100em}
\end{figure}

As mentioned in Section \ref{sec:Modelling a typical cooling system}, the VGR uses its suction cup to transport the workpiece from the SL to the rack feeder of the AHBW. The solenoid present in the suction cup is turned on/off if the position of the horizontal and vertical axis of the VGR is above or below a threshold position. The threshold position is a measurement value (i.e., $y_k$) and can be estimated by \textit{BayesImposter}. The correct value of the threshold position where the suction cup is turned off (release the workpiece) is 2 cm. The estimated value of the threshold position is also calculated as 2 cm using \textit{BayesImposter} at a particular state (i.e., moving from SL to AHBW). After the successful estimation of the threshold position with all other tag values of the \textit{victim page} using the same \textit{BayesImposter}, the attacker can recreate the .bss imposter page. Now, the attacker initiates the memory deduplication + Rowhammer attack and arbitrarily causes a bit-flip in the .bss imposter page. A demonstration of the attack is shown in Fig. \ref{fig:rowhammer_bit_flip}, which indicates the location of the occurred bit-flip in the victim row.  \textit{(0 0 1 7 3c97 0)} means address of channel 0, dimm 0, rank 1, bank 7, row 3c97, column 0 in DRAM with a row-offset 0743, which has a byte value \textit{f7} after the bit-flip; however, byte expected according to fill pattern is \textit{ff (i.e., all erased)}. The victim byte \textit{f7} is the upper byte of the threshold position being corrupted that changes the 2 cm threshold position to 2050 cm. This causes an out-of-range value for the VGR resulting in a wrong drop-off location of the workpiece other than the rack-feeder. This may result in possible equipment damage or even can kill a person if the attacker drops the workpiece on a target person. A video demonstration of this attack is given here: {\color{blue}\url{https://sites.google.com/view/bayesmem/home}}

\vspace{-00.8500em}
\subsection {\textbf{Adversarial control using BayesImposter}}
\label{subsec:Adversarial Control}
\begin{figure}[ht!]
 \vspace{-1.50em}
\centering
\includegraphics[width=0.4\textwidth,height=0.15\textheight]{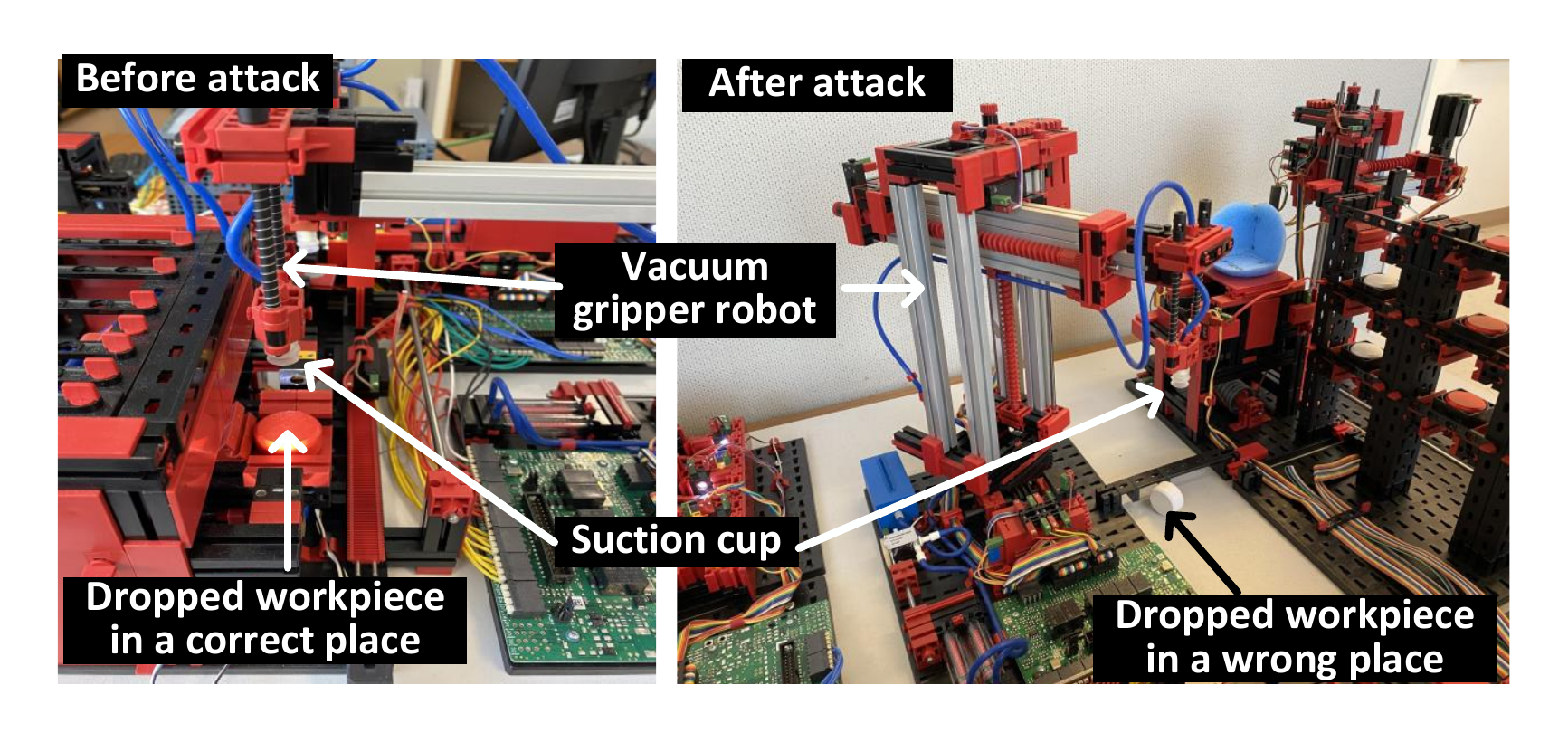}
\vspace{-1.930em}
\caption{Dropping workpiece using adversarial control.}
\label{fig:adversarial}
\vspace{-01.2800em}
\end{figure}

As the attacker knows the physical location of a tag value in the tag table of the .bss imposter page, he can target a particular tag value and initiate an adversarial control over that tag value. For example, the attacker can cause a bit-flip of  \texttt{suctionstate} from $1 \rightarrow 0$ and can adversarially drop the workpiece from the suction cup when it is not supposed to drop the workpiece ( Fig. \ref{fig:adversarial}). This may result in possible equipment damage or even can kill a person if the attacker drops the workpiece on a target person. This adversarial control makes \textit{BayesImposter} stronger compared to ~\cite{razavi2016flip, barresi2015cain, bosman2016dedup,oliverio2017secure}.

 \vspace{-0.70em}
\subsection {\textbf{Profiling time in our testbed}}
\label{subsec:Profiling time in our testbed}

\begin{figure}[ht!]
 \vspace{-1.50em}
\centering
\includegraphics[width=0.3\textwidth,height=0.14\textheight]{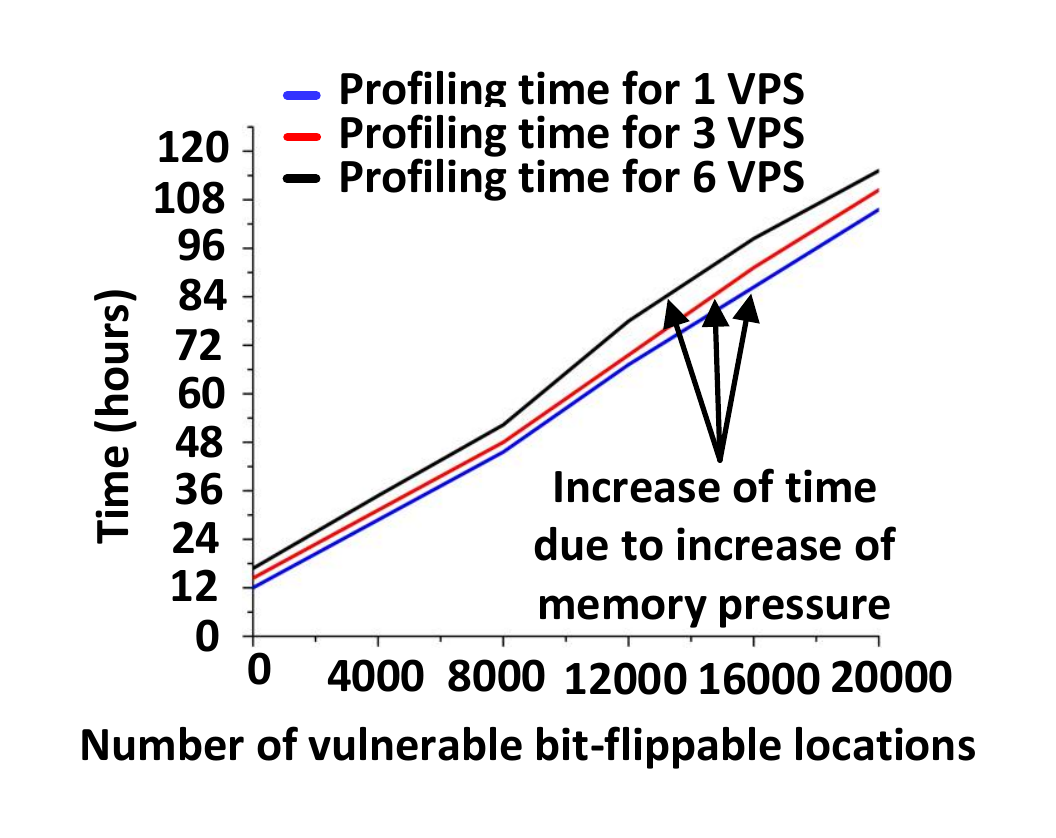}
\vspace{-1.540em}
\caption{Profiling time for different number of VPSs.}
\label{fig:profilingtime}
\vspace{-0.100em}
\end{figure}

Fig. \ref{fig:profilingtime} evaluates the profiling time (see Section \ref{subsec:Merging the duplicated .bss section}) for different number of VPSs in the cloud. \textit{BayesImposter} takes $\sim$51.45 seconds to complete single-sided Rowhammer for each target row. We searched for vulnerable locations for the Rowhammer in the memory space, and Fig. \ref{fig:profilingtime} shows that to get $\sim$20000 vulnerable locations,  $\sim$100 hours are required. With the increase of VPSs, this profiling time increases due to more memory pressure in the system memory. Fig. \ref{fig:profilingtime} shows the profiling time for 1, 3, and 6 VPSs in the same cloud.

\vspace{-01.100em}
\subsection {\textbf{Attack time}}
\label{subsec:Attack time}

Here, we define attack time as how much time it takes to cause a bit flip in the .bss section. Attack time is the summation of the memory deduplication time and the Rowhammer implementation time. The exact time required for memory deduplication can be calculated using the timing side-channel \cite{bosman2016dedup}. However, roughly, the maximum time for memory deduplication is the time needed to scan all the memory of the co-located VPSs in the cloud. Here, for simplicity, we assume that deduplication happens within this maximum time frame, and hence, we consider this maximum time as the memory deduplication time. The memory deduplication time depends upon the parameters \texttt{pages\_to\_scan} and \texttt{sleep
\_millisec}. In default configuration, \texttt{pages\_to\_scan} = 100 and \texttt{sleep
\_millisec} = 20. Therefore, Linux/KSM can scan 1000 pages/second, which results in a total scan time of almost 5 minutes per 1GB of main memory \cite{miller2013xlh}.  As the victim VPS has a main memory of 2 GB, it should take approximately 10 minutes to scan all the pages in the main memory of a VPS. In our testbed, the memory deduplication takes approx. 13 minutes, and the Rowhammering process takes approx. 51.45 seconds to complete a single-sided Rowhammer for each target row. Therefore, after summing up these two figures, the total attack time is approximately 13 minutes and 52 seconds for 1 target VPS.

\begin{figure}[ht!]
\vspace{-1.310em}
\centering
\includegraphics[width=0.3\textwidth,height=0.14\textheight]{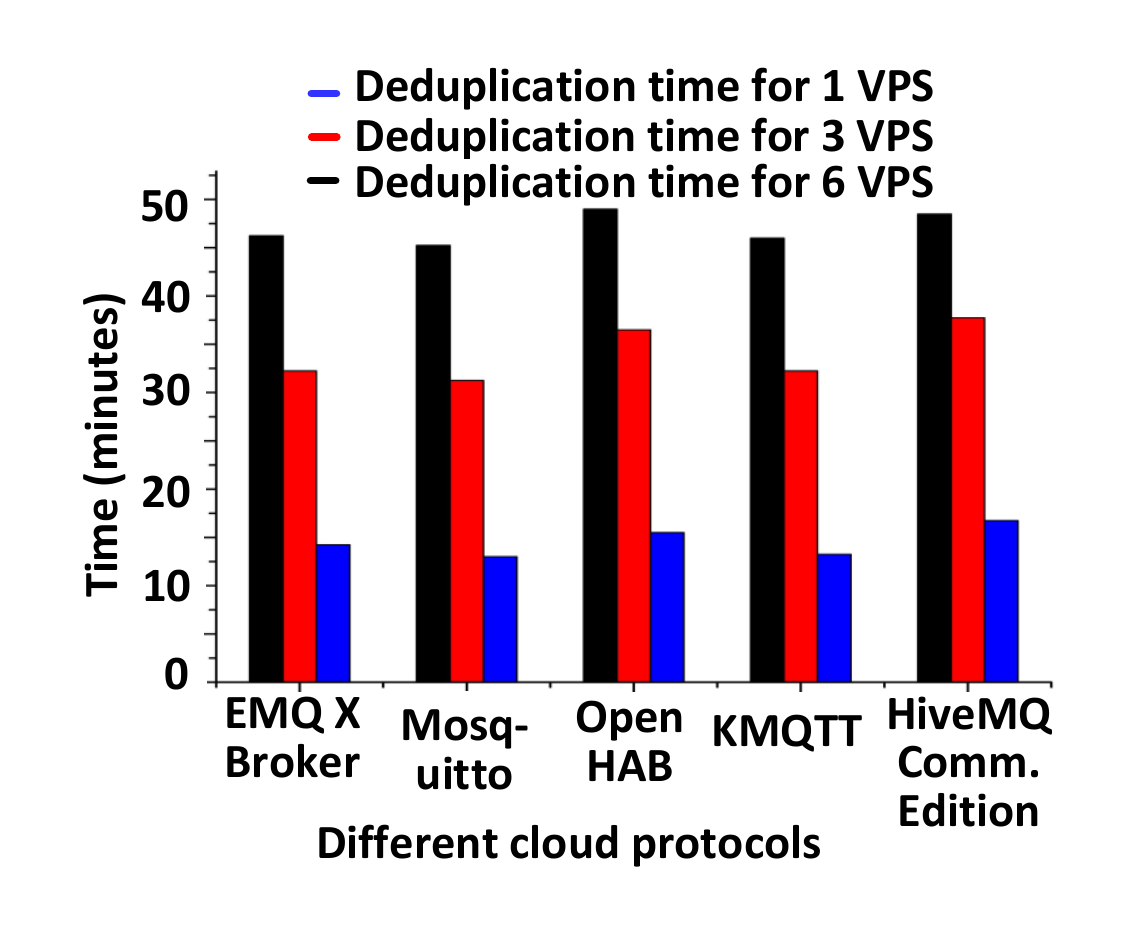}
\vspace{-1.90em}
\caption{Deduplication time for different protocols.}
\label{fig:deduptime_protocols}
\vspace{-01.210em}
\end{figure}

Fig. \ref{fig:deduptime_protocols} shows the memory deduplication time for five variants of MQTT cloud protocol for 1, 3, and 6 VPSs. This figure indicates that all five variants of the cloud protocol give almost equal deduplication time. As the addition of a VPS increases the scannable memory locations, the deduplication time increases with  the number of co-located VPS in the cloud. The Rowhammer implementation time for a target row is almost the same for all five protocol variants.  

\vspace{-.500em}
\subsection {\textbf{Evaluation for different cloud protocols}}
\label{subsec:Attacking the stateThermostat}

As our attack model does not require any software bug present in the implementation of cloud protocols, state-of-the-art variants of cloud protocols should be vulnerable to our attack model. To support this claim, we implement a total of five variants of the MQTT protocol in our testbed and find that all are equally vulnerable, which proves the generalization of our attack model in ICSs.

\begin{table}[h!]
\vspace{-0.700em}
	\footnotesize
	\centering
		\caption{Cloud protocol variants vulnerable to \textit{BayesImposter}}
		\vspace{-1.40em}
		\label{table:different cloud protocols}
		\begin{tabular}{|p{0.4cm}|p{3.350cm}|p{1.8cm}|}
			\hline
			\cellcolor [gray]{0.85} Sl. &  \cellcolor [gray]{0.85} Cloud protocol variants &  \cellcolor [gray]{0.85} Vulnerability \\
			\hline
			\hline
    		1 & EMQ X  Broker \cite{EMQXbroker}& \checkmark\\
			\hline
			2 & Mosquitto \cite{mosquitto}  & \checkmark\\
			\hline
			3 & MQTT-C  \cite{mqttc} & \checkmark\\
			\hline
			4 & eMQTT5 \cite{eMQTT} & \checkmark\\
			\hline
			5 & wolfMQTT  \cite{wolfmqtt} & \checkmark\\
			\hline
		\end{tabular}
	\vspace{-0.5em}
\end{table}
\vspace{-0.5em}




\section{Defense}
\label{sec:defense}

The following mitigations should be adopted against \textit{BayesImposter}.

\vspace{0.0em}
\textbf{Increasing entropy in the .bss section:} 
To prevent the attack, we increase entropy in the .bss section. This is done using a random variable as a signature in the .bss section. 
The attacker requires a significant amount of memory and time to break this signature variable \cite{barresi2015cain} as this variable is not a part of the state variable. 
This approach is also effective against a malicious insider.

\vspace{0.0em}
\textbf{Securing cloud server from the malicious VPS:} 
Any unauthorized cloud provider or personnel, or visitor should not access the cloud server without the presence of authorized personnel. Periodic screening by an authorized person needs to be carried out to look for any unauthorized co-hosted VPS. Any unnecessary or suspicious co-located VPS should be considered as a security breach and should be immediately contained in the cloud.

\vspace{0.0em}
\textbf{Turning off the KSM:} 
To prevent memory deduplication, KSM can be turned permanently off. KSM is off by default in recent Linux kernel \cite{KSM2000fedora}. However, the KSM service, which is included in the \texttt{qemu-kvm} package, is turned on by the KVM host in the cloud setting. We turn off the KSM using the \texttt{ksm/ksmtuned} services in the KVM host. However, turning off the KSM may increase memory usage in clouds. Therefore, it is not favorable where memory workloads are high in cloud settings \cite{jia2015coordinate}. 

\vspace{0.0em}
\textbf{Preventing Rowhammer in DRAM:} The next way to prevent \textit{BayesImposter} is to prevent the Rowhammer in DRAM. While the built-in error-correcting codes (ECCs) can prevent single bit-flip in  64-bit words \cite{cojocar2019exploiting}, it may not be enough where the Rowhammer causes multiple bit-flips \cite{aichinger2015ddr, lanteigne2016rowhammer}. While only modern AMD Ryzen processors support ECC RAM in consumer hardware, Intel restricts its support to server CPUs \cite{8418607}. One method to prevent Rowhammer is to increase (e.g., double) the refresh rate in DRAM chips \cite{mutlu2017rowhammer}. This can reduce the probability of multiple bit-flips in DRAM, but causes more energy consumption and more overhead in the memory  \cite{emma2008rethinking, kim2014flipping}. Another method is to probabilistically open adjacent or non-adjacent rows, whenever a row is opened or closed \cite{kim2014architectural}. An introduction of a redundant array of independent memory (i.e., RAIM) \cite{meaney2012ibm}, and ANVIL \cite{aweke2016anvil} in the server hardware can make the Rowhammer attack infeasible. 
Moreover, replacing  older chips with DDR4 having Target Row Refresh (TRR) capability can prevent single-sided and multi-sided  Rowhammer attack on cloud networks \cite{kwong2020rambleed}. However,  \cite {frigo2020trrespass} shows that  DDR4 can also be compromised using TRR-aware attacks.

\vspace{-0.5000em}
\section{Related Work}
\label{sec:Related Work}


\vspace{0.0em}
\noindent \textbf{Attacks on ICSs:} The attacks on ICSs can be broadly classified  as attacks on physical hardware (e.g., PLCs, control modules, etc.),  attacks on communication networks, and attacks on  sensing side.

Abbasi et al. \cite{abbasi2016ghost} demonstrated an attack on PLCs by exploiting pin control  operations of certain input/output pins resulting in abnormal hardware interrupt in PLCs. Garcia et al. \cite{garcia2017hey} presented a malware-PLC rootkit that can attack PLCs using the physics of the underlying systems. Bolshev et al. \cite{bolshev2016rising} showed an attack on the physical layer (i.e., analog-to-digital converter), resulting in false data injection into PLCs. Spenneberg et al. \cite{spenneberg2016plc} developed a worm - PLC Blaster, that independently searches any network for S7-1200v3 devices and attacks them when the protective mechanisms are switched off. \textit{Compared to our attack model, these attacks on PLCs lack the presence of adversarial control over PLCs and do not provide any means of stealthiness with respect to the monitoring entity.} 

Klick et al. \cite{klick2015internet} showed that internet-facing controllers act as an SNMP scanner or SOCKS proxy, and their protocols can be misused by an adversary to inject false codes into PLCs, which are not directly connected to the internet. Basnight et al. \cite{basnight2013firmware} presented an attack on firmware exploiting communication protocols of PLCs. Beresford et al. \cite{beresford2011exploiting} discovered vulnerabilities in Siemens S7 series communication protocol and showed a replay attack on ICSs. \textit{Compared to these attacks, our attack model does not need any vulnerabilities in the communication protocol and does work without any presence of software bugs at any level of the system.}  

Barua et al. \cite{barua2020hall,barua2022pressuresensor,Barua2019,barua2020special,barua2022premsat}, Liu et al. \cite{liu2011false}, and McLaughlin et al. \cite{mclaughlin2014controller} showed \textit{false data injection} attack on different sensing nodes of ICSs leading to abnormal behaviour of the underlying system. \textit{Compared to these attacks, our attack model is capable of false command injection from a remote location with adversarial control in ICSs.}

\vspace{0.2em}
\noindent \textbf{Attacks using memory deduplication and/or Rowhammer:}  Bosman et al. \cite{bosman2016dedup} demonstrated memory deduplication based exploitation vector on Windows using Microsoft Edge. 
Barresi et al. \cite{barresi2015cain} exploited the memory deduplication in a virtualized environment to break ASLR of Windows and Linux. This attack uses brute force to duplicate the target page in the memory. Razavi et al. \cite{razavi2016flip} provided Flip  Fleng Shui (FFS) to break cryptosystems using both the memory deduplication and Rowhammer. \textbf{There are fundamental differences between our work and  \cite{bosman2016dedup,barresi2015cain,razavi2016flip}.} \textit{\textbf{First}, our attack model exploited the .bss section of cloud protocols that is more impactful and realistic in ICSs. \textbf{Second}, our attack uses the Bayesian estimation to duplicate the target page compared to the brute force approach in \cite{bosman2016dedup,barresi2015cain,razavi2016flip}. This results in significantly less memory usage (i.e., in KB compared to GB) and time (i.e., in minutes compared to hours) to duplicate the target page. This makes our attack model more feasible. \textbf{Third,} our attack model demonstrates adversarial control over the target ICS that is absent in \cite{bosman2016dedup,barresi2015cain,razavi2016flip}. }

Seaborn et al. \cite{seaborn2015exploiting} exploited CPU caches to read directly from DRAM using the Rowhammer bug. Gruss et al. \cite{gruss2016rowhammer} used cache eviction sets and Transparent Huge Pages (THP) for a successful double-sided Rowhammer. Tatar et al. \cite{tatar2018throwhammer} used Rowhammer attacks over the network to cause bit-flips using Remote DMA (RDMA). \textit{Compared to these works, our work uses memory deduplication to skip the knowledge of physical memory location and uses single-sided Rowhammer on the target cloud memory. Moreover, our attack does not require any RDMA to happen that makes our attack more flexible in the context of ICSs.}

\vspace{-0em}
\section{Conclusion}
\label{sec:conclusion}

We present an attack model-\textit{BayesImposter} that can hamper the availability and integrity of an ICS in cloud settings.  We are the first to point out how the .bss section of the target control DLL file of cloud protocols is vulnerable in ICS. \textit{BayesImposter} exploits the memory deduplication feature of the cloud that merges the attacker's provided .bss imposter page with the victim page. To create the .bss imposter page, \textit{BayesImposter} uses a new technique that involves the \textit{Bayesian estimation}, which results in less memory and time compared to recent works \cite{bosman2016dedup,barresi2015cain,razavi2016flip}. We show that as ICSs can be expressed as state-space models; hence, the \textit{Bayesian estimation} is an ideal choice to be combined with the memory deduplication in cloud settings. We prepare a scaled-down model of an automated high-bay warehouse using SIMATIC PLC from Siemens and demonstrate our attack model on this practical testbed. We show that our attack model is effective on different variants of cloud protocols, and does not need any vulnerabilities in the cloud protocol, and  works without any presence of software bug in any level of the system that proves a generalization of our attack model. We show that \textit{BayesImposter} is capable of \textit{adversarial control} that can cause severe consequences through \textit{system demage}. Therefore, our attack is impactful, and the countermeasures should be adopted to prevent any future attack like ours in ICSs.

\begin{acks}

This work was partially supported by the National Science Foundation (NSF) under awards CMMI-1739503 and ECCS-2028269. Any opinions, findings, conclusions, or recommendations expressed in this paper are those of the authors and do not necessarily reflect the views of the funding agencies.

\end{acks}

\bibliographystyle{ACM-Reference-Format}
\bibliography{bibfile}


\begin{thebibliography}{75}


\ifx \showCODEN    \undefined \def \showCODEN     #1{\unskip}     \fi
\ifx \showDOI      \undefined \def \showDOI       #1{#1}\fi
\ifx \showISBNx    \undefined \def \showISBNx     #1{\unskip}     \fi
\ifx \showISBNxiii \undefined \def \showISBNxiii  #1{\unskip}     \fi
\ifx \showISSN     \undefined \def \showISSN      #1{\unskip}     \fi
\ifx \showLCCN     \undefined \def \showLCCN      #1{\unskip}     \fi
\ifx \shownote     \undefined \def \shownote      #1{#1}          \fi
\ifx \showarticletitle \undefined \def \showarticletitle #1{#1}   \fi
\ifx \showURL      \undefined \def \showURL       {\relax}        \fi
\providecommand\bibfield[2]{#2}
\providecommand\bibinfo[2]{#2}
\providecommand\natexlab[1]{#1}
\providecommand\showeprint[2][]{arXiv:#2}

\bibitem[com({[n.\,d.]})]%
        {commerciallicence}
 \bibinfo{year}{[n.\,d.]}\natexlab{}.
\newblock \bibinfo{booktitle}{\emph{{Affordable MQTT Broker Pricing}}}.
\newblock
\newblock
\shownote{\url{https://www.bevywise.com/mqtt-broker/pricing.html}}.


\bibitem[KSM({[n.\,d.]})]%
        {KSM2000fedora}
 \bibinfo{year}{[n.\,d.]}\natexlab{}.
\newblock \bibinfo{booktitle}{\emph{{Chapter 7. KSM}}}.
\newblock
\newblock
\shownote{\url{https://docs.fedoraproject.org/en-US/Fedora/18/html/Virtualization_Administration_Guide/chap-KSM.html}}.


\bibitem[dat({[n.\,d.]})]%
        {datadedup}
 \bibinfo{year}{[n.\,d.]}\natexlab{}.
\newblock \bibinfo{booktitle}{\emph{{Data Deduplication Overview}}}.
\newblock
\newblock
\shownote{\url{https://docs.microsoft.com/en-us/windows-server/storage/data-deduplication/overview}}.


\bibitem[EMQ({[n.\,d.]})]%
        {EMQXbroker}
 \bibinfo{year}{[n.\,d.]}\natexlab{}.
\newblock \bibinfo{booktitle}{\emph{{EMQ X Broker}}}.
\newblock
\newblock
\shownote{\url{https://www.emqx.io/downloads\#broker}}.


\bibitem[eMQ({[n.\,d.]})]%
        {eMQTT}
 \bibinfo{year}{[n.\,d.]}\natexlab{}.
\newblock \bibinfo{booktitle}{\emph{{eMQTT5}}}.
\newblock
\newblock
\shownote{\url{https://github.com/X-Ryl669/eMQTT5}}.


\bibitem[Fac({[n.\,d.]})]%
        {FactorySimulation}
 \bibinfo{year}{[n.\,d.]}\natexlab{}.
\newblock \bibinfo{booktitle}{\emph{{Factory Simulation 24V}}}.
\newblock
\newblock
\shownote{\url{https://www.fischertechnik.de/en/service/elearning/simulating/fabrik-simulation-24v}.
  (Accessed: 03-22-2022)}.


\bibitem[TIA({[n.\,d.]})]%
        {TIAportal}
 \bibinfo{year}{[n.\,d.]}\natexlab{}.
\newblock \bibinfo{booktitle}{\emph{{How to use TIA Portal Cloud}}}.
\newblock
\newblock
\shownote{\url{https://new.siemens.com/global/en/products/automation/industry-software/automation-software/tia-portal/highlights/tia-portal-cloud.html}}.


\bibitem[kSM({[n.\,d.]})]%
        {kSM1000linux}
 \bibinfo{year}{[n.\,d.]}\natexlab{}.
\newblock \bibinfo{booktitle}{\emph{{Linux kernel 2.6.32, Section 1.3. Kernel
  Samepage Merging (memory deduplication)}}}.
\newblock
\newblock
\shownote{\url{https://kernelnewbies.org/Linux_2_6_32##Kernel_Samepage_Merging_.28memory_deduplication.29}}.


\bibitem[mos({[n.\,d.]})]%
        {mosquitto}
 \bibinfo{year}{[n.\,d.]}\natexlab{}.
\newblock \bibinfo{booktitle}{\emph{{mosquitto}}}.
\newblock
\newblock
\shownote{\url{https://mosquitto.org/}}.


\bibitem[mqt({[n.\,d.]})]%
        {mqttc}
 \bibinfo{year}{[n.\,d.]}\natexlab{}.
\newblock \bibinfo{booktitle}{\emph{{MQTT-C}}}.
\newblock
\newblock
\shownote{\url{https://github.com/LiamBindle/MQTT-C}}.


\bibitem[sie({[n.\,d.]})]%
        {siemensVPS}
 \bibinfo{year}{[n.\,d.]}\natexlab{}.
\newblock \bibinfo{booktitle}{\emph{{Siemens: How to connect to a PLC with TIA
  Portal in a Virtual Machine}}}.
\newblock
\newblock
\shownote{\url{https://web.awc-inc.com/siemens-how-to-connect-to-a-plc-with-tia-portal-in-a-virtual-machine/}}.


\bibitem[SIM({[n.\,d.]})]%
        {SIMATIC}
 \bibinfo{year}{[n.\,d.]}\natexlab{}.
\newblock \bibinfo{booktitle}{\emph{{SIMATIC S7-1500}}}.
\newblock
\newblock
\shownote{\url{https://cache.industry.siemens.com/dl/files/914/59191914/att_86487/v1/s71500_cpu1516_3_pn_dp_manual_en-US_en-US.pdf}}.


\bibitem[wol({[n.\,d.]})]%
        {wolfmqtt}
 \bibinfo{year}{[n.\,d.]}\natexlab{}.
\newblock \bibinfo{booktitle}{\emph{{wolfMQTT}}}.
\newblock
\newblock
\shownote{\url{https://github.com/wolfSSL/wolfMQTT}}.


\bibitem[Abbasi and Hashemi(2016)]%
        {abbasi2016ghost}
\bibfield{author}{\bibinfo{person}{Ali Abbasi} {and} \bibinfo{person}{Majid
  Hashemi}.} \bibinfo{year}{2016}\natexlab{}.
\newblock \showarticletitle{Ghost in the plc designing an undetectable
  programmable logic controller rootkit via pin control attack}.
\newblock \bibinfo{journal}{\emph{Black Hat Europe}}  \bibinfo{volume}{2016}
  (\bibinfo{year}{2016}), \bibinfo{pages}{1--35}.
\newblock


\bibitem[Aichinger(2015)]%
        {aichinger2015ddr}
\bibfield{author}{\bibinfo{person}{Barbara Aichinger}.}
  \bibinfo{year}{2015}\natexlab{}.
\newblock \showarticletitle{DDR memory errors caused by Row Hammer}. In
  \bibinfo{booktitle}{\emph{2015 IEEE High Performance Extreme Computing
  Conference (HPEC)}}. IEEE, \bibinfo{pages}{1--5}.
\newblock


\bibitem[Annapoorani et~al\mbox{.}(2018)]%
        {annapoorani2018analysis}
\bibfield{author}{\bibinfo{person}{S Annapoorani}, \bibinfo{person}{B
  Srinivasan}, {and} \bibinfo{person}{GA Mylavathi}.}
  \bibinfo{year}{2018}\natexlab{}.
\newblock \showarticletitle{Analysis of various virtual machine attacks in
  cloud computing}. In \bibinfo{booktitle}{\emph{2018 2nd international
  Conference on Inventive Systems and Control (ICISC)}}. IEEE,
  \bibinfo{pages}{1016--1019}.
\newblock


\bibitem[Appelbaum et~al\mbox{.}(2013)]%
        {spiegel}
\bibfield{author}{\bibinfo{person}{J.R. Appelbaum}, \bibinfo{person}{L.
  Poitras}, \bibinfo{person}{M. Rosenbach}, \bibinfo{person}{C. St{\"o}cker},
  \bibinfo{person}{J. Schindler}, {and} \bibinfo{person}{H. Stark}.}
  \bibinfo{year}{2013}\natexlab{}.
\newblock \showarticletitle{{Inside TAO : documents reveal top NSA hacking
  unit}}.
\newblock \bibinfo{journal}{\emph{Der Spiegel}} (\bibinfo{date}{29 12}
  \bibinfo{year}{2013}).
\newblock
\showISSN{0038-7452}


\bibitem[Aweke et~al\mbox{.}(2016)]%
        {aweke2016anvil}
\bibfield{author}{\bibinfo{person}{Zelalem~Birhanu Aweke},
  \bibinfo{person}{Salessawi~Ferede Yitbarek}, \bibinfo{person}{Rui Qiao},
  \bibinfo{person}{Reetuparna Das}, \bibinfo{person}{Matthew Hicks},
  \bibinfo{person}{Yossi Oren}, {and} \bibinfo{person}{Todd Austin}.}
  \bibinfo{year}{2016}\natexlab{}.
\newblock \showarticletitle{ANVIL: Software-based protection against
  next-generation rowhammer attacks}.
\newblock \bibinfo{journal}{\emph{ACM SIGPLAN Notices}} \bibinfo{volume}{51},
  \bibinfo{number}{4} (\bibinfo{year}{2016}), \bibinfo{pages}{743--755}.
\newblock


\bibitem[Barresi et~al\mbox{.}(2015)]%
        {barresi2015cain}
\bibfield{author}{\bibinfo{person}{Antonio Barresi}, \bibinfo{person}{Kaveh
  Razavi}, \bibinfo{person}{Mathias Payer}, {and} \bibinfo{person}{Thomas~R
  Gross}.} \bibinfo{year}{2015}\natexlab{}.
\newblock \showarticletitle{$\{$CAIN$\}$: Silently Breaking $\{$ASLR$\}$ in the
  Cloud}. In \bibinfo{booktitle}{\emph{9th $\{$USENIX$\}$ Workshop on Offensive
  Technologies ($\{$WOOT$\}$ 15)}}.
\newblock


\bibitem[Bartodziej(2017)]%
        {bartodziej2017concept}
\bibfield{author}{\bibinfo{person}{Christoph~Jan Bartodziej}.}
  \bibinfo{year}{2017}\natexlab{}.
\newblock \showarticletitle{The concept industry 4.0}.
\newblock In \bibinfo{booktitle}{\emph{The concept industry 4.0}}.
  \bibinfo{publisher}{Springer}, \bibinfo{pages}{27--50}.
\newblock


\bibitem[Barua and Al~Faruque(2019)]%
        {Barua2019}
\bibfield{author}{\bibinfo{person}{Anomadarshi Barua} {and}
  \bibinfo{person}{Mohammad~Abdullah Al~Faruque}.}
  \bibinfo{year}{2019}\natexlab{}.
\newblock \bibinfo{booktitle}{\emph{The Hall Sensor Security}}.
\newblock \bibinfo{publisher}{Springer Berlin Heidelberg},
  \bibinfo{address}{Berlin, Heidelberg}, \bibinfo{pages}{1--4}.
\newblock
\showISBNx{978-3-642-27739-9}
\urldef\tempurl%
\url{https://doi.org/10.1007/978-3-642-27739-9_1652-1}
\showDOI{\tempurl}


\bibitem[Barua and Al~Faruque(2020a)]%
        {barua2020hall}
\bibfield{author}{\bibinfo{person}{Anomadarshi Barua} {and}
  \bibinfo{person}{Mohammad~Abdullah Al~Faruque}.}
  \bibinfo{year}{2020}\natexlab{a}.
\newblock \showarticletitle{Hall Spoofing: A Non-Invasive DoS Attack on
  Grid-Tied Solar Inverter}. In \bibinfo{booktitle}{\emph{29th $\{$USENIX$\}$
  Security Symposium ($\{$USENIX$\}$ Security 20)}}.
  \bibinfo{pages}{1273--1290}.
\newblock


\bibitem[Barua and Al~Faruque(2020b)]%
        {barua2020special}
\bibfield{author}{\bibinfo{person}{Anomadarshi Barua} {and}
  \bibinfo{person}{Mohammad~Abdullah Al~Faruque}.}
  \bibinfo{year}{2020}\natexlab{b}.
\newblock \showarticletitle{Special session: Noninvasive sensor-spoofing
  attacks on embedded and cyber-physical systems}. In
  \bibinfo{booktitle}{\emph{2020 IEEE 38th International Conference on Computer
  Design (ICCD)}}. IEEE, \bibinfo{pages}{45--48}.
\newblock


\bibitem[Barua and Al~Faruque(2022a)]%
        {barua2022pressuresensor}
\bibfield{author}{\bibinfo{person}{Anomadarshi Barua} {and}
  \bibinfo{person}{Mohammad~Abdullah Al~Faruque}.}
  \bibinfo{year}{2022}\natexlab{a}.
\newblock \showarticletitle{{A Wolf in Sheep’s Clothing: Spreading Deadly
  Pathogens Under the Disguise of Popular Music}}. In
  \bibinfo{booktitle}{\emph{29th ACM Conference on Computer and Communications
  Security (CCS)}}.
\newblock


\bibitem[Barua and Al~Faruque(2022b)]%
        {barua2022premsat}
\bibfield{author}{\bibinfo{person}{Anomadarshi Barua} {and}
  \bibinfo{person}{Mohammad~Abdullah Al~Faruque}.}
  \bibinfo{year}{2022}\natexlab{b}.
\newblock \showarticletitle{{PreMSat: Preventing Magnetic Saturation Attack on
  Hall Sensors}}. In \bibinfo{booktitle}{\emph{International Conference on
  Cryptographic Hardware and Embedded Systems (TCHES 2022)}}.
\newblock


\bibitem[Basnight et~al\mbox{.}(2013)]%
        {basnight2013firmware}
\bibfield{author}{\bibinfo{person}{Zachry Basnight}, \bibinfo{person}{Jonathan
  Butts}, \bibinfo{person}{Juan Lopez~Jr}, {and} \bibinfo{person}{Thomas
  Dube}.} \bibinfo{year}{2013}\natexlab{}.
\newblock \showarticletitle{Firmware modification attacks on programmable logic
  controllers}.
\newblock \bibinfo{journal}{\emph{International Journal of Critical
  Infrastructure Protection}} \bibinfo{volume}{6}, \bibinfo{number}{2}
  (\bibinfo{year}{2013}), \bibinfo{pages}{76--84}.
\newblock


\bibitem[Beresford(2011)]%
        {beresford2011exploiting}
\bibfield{author}{\bibinfo{person}{Dillon Beresford}.}
  \bibinfo{year}{2011}\natexlab{}.
\newblock \showarticletitle{Exploiting siemens simatic s7 plcs}.
\newblock \bibinfo{journal}{\emph{Black Hat USA}} \bibinfo{volume}{16},
  \bibinfo{number}{2} (\bibinfo{year}{2011}), \bibinfo{pages}{723--733}.
\newblock


\bibitem[Bolshev et~al\mbox{.}(2016)]%
        {bolshev2016rising}
\bibfield{author}{\bibinfo{person}{Alexander Bolshev}, \bibinfo{person}{Jason
  Larsen}, \bibinfo{person}{Marina Krotofil}, {and} \bibinfo{person}{Reid
  Wightman}.} \bibinfo{year}{2016}\natexlab{}.
\newblock \showarticletitle{A rising tide: Design exploits in industrial
  control systems}. In \bibinfo{booktitle}{\emph{10th $\{$USENIX$\}$ Workshop
  on Offensive Technologies ($\{$WOOT$\}$ 16)}}.
\newblock


\bibitem[Bosman et~al\mbox{.}(2016)]%
        {bosman2016dedup}
\bibfield{author}{\bibinfo{person}{Erik Bosman}, \bibinfo{person}{Kaveh
  Razavi}, \bibinfo{person}{Herbert Bos}, {and} \bibinfo{person}{Cristiano
  Giuffrida}.} \bibinfo{year}{2016}\natexlab{}.
\newblock \showarticletitle{Dedup est machina: Memory deduplication as an
  advanced exploitation vector}. In \bibinfo{booktitle}{\emph{2016 IEEE
  symposium on security and privacy (SP)}}. IEEE, \bibinfo{pages}{987--1004}.
\newblock


\bibitem[Chang et~al\mbox{.}(2011)]%
        {chang2011empirical}
\bibfield{author}{\bibinfo{person}{Chao-Rui Chang}, \bibinfo{person}{Jan-Jan
  Wu}, {and} \bibinfo{person}{Pangfeng Liu}.} \bibinfo{year}{2011}\natexlab{}.
\newblock \showarticletitle{An empirical study on memory sharing of virtual
  machines for server consolidation}. In \bibinfo{booktitle}{\emph{2011 IEEE
  Ninth International Symposium on Parallel and Distributed Processing with
  Applications}}. IEEE, \bibinfo{pages}{244--249}.
\newblock


\bibitem[Choi et~al\mbox{.}(2020)]%
        {choi2020expansion}
\bibfield{author}{\bibinfo{person}{Seungoh Choi}, \bibinfo{person}{Jongwon
  Choi}, \bibinfo{person}{Jeong-Han Yun}, \bibinfo{person}{Byung-Gil Min},
  {and} \bibinfo{person}{HyoungChun Kim}.} \bibinfo{year}{2020}\natexlab{}.
\newblock \showarticletitle{Expansion of $\{$ICS$\}$ testbed for security
  validation based on $\{$MITRE$\}$ atT\&Ck techniques}. In
  \bibinfo{booktitle}{\emph{13th $\{$USENIX$\}$ Workshop on Cyber Security
  Experimentation and Test ($\{$CSET$\}$ 20)}}.
\newblock


\bibitem[Cojocar et~al\mbox{.}(2019)]%
        {cojocar2019exploiting}
\bibfield{author}{\bibinfo{person}{Lucian Cojocar}, \bibinfo{person}{Kaveh
  Razavi}, \bibinfo{person}{Cristiano Giuffrida}, {and}
  \bibinfo{person}{Herbert Bos}.} \bibinfo{year}{2019}\natexlab{}.
\newblock \showarticletitle{Exploiting correcting codes: On the effectiveness
  of ECC memory against Rowhammer attacks}. In \bibinfo{booktitle}{\emph{2019
  IEEE Symposium on Security and Privacy (SP)}}. IEEE, \bibinfo{pages}{55--71}.
\newblock


\bibitem[Deng et~al\mbox{.}(2017)]%
        {deng2017memory}
\bibfield{author}{\bibinfo{person}{Yuhui Deng}, \bibinfo{person}{Xinyu Huang},
  \bibinfo{person}{Liangshan Song}, \bibinfo{person}{Yongtao Zhou}, {and}
  \bibinfo{person}{Frank~Z Wang}.} \bibinfo{year}{2017}\natexlab{}.
\newblock \showarticletitle{Memory deduplication: An effective approach to
  improve the memory system}.
\newblock \bibinfo{journal}{\emph{Journal of Information Science and
  Engineering}} \bibinfo{volume}{33}, \bibinfo{number}{5}
  (\bibinfo{year}{2017}), \bibinfo{pages}{1103--1120}.
\newblock


\bibitem[Emma et~al\mbox{.}(2008)]%
        {emma2008rethinking}
\bibfield{author}{\bibinfo{person}{Philip~G Emma}, \bibinfo{person}{William~R
  Reohr}, {and} \bibinfo{person}{Mesut Meterelliyoz}.}
  \bibinfo{year}{2008}\natexlab{}.
\newblock \showarticletitle{Rethinking refresh: Increasing availability and
  reducing power in DRAM for cache applications}.
\newblock \bibinfo{journal}{\emph{IEEE micro}} \bibinfo{volume}{28},
  \bibinfo{number}{6} (\bibinfo{year}{2008}), \bibinfo{pages}{47--56}.
\newblock


\bibitem[Friedland(2012)]%
        {friedland2012control}
\bibfield{author}{\bibinfo{person}{Bernard Friedland}.}
  \bibinfo{year}{2012}\natexlab{}.
\newblock \bibinfo{booktitle}{\emph{Control system design: an introduction to
  state-space methods}}.
\newblock \bibinfo{publisher}{Courier Corporation}.
\newblock


\bibitem[Frigo et~al\mbox{.}(2020)]%
        {frigo2020trrespass}
\bibfield{author}{\bibinfo{person}{Pietro Frigo}, \bibinfo{person}{Emanuele
  Vannacc}, \bibinfo{person}{Hasan Hassan}, \bibinfo{person}{Victor Van
  Der~Veen}, \bibinfo{person}{Onur Mutlu}, \bibinfo{person}{Cristiano
  Giuffrida}, \bibinfo{person}{Herbert Bos}, {and} \bibinfo{person}{Kaveh
  Razavi}.} \bibinfo{year}{2020}\natexlab{}.
\newblock \showarticletitle{TRRespass: Exploiting the many sides of target row
  refresh}. In \bibinfo{booktitle}{\emph{2020 IEEE Symposium on Security and
  Privacy (SP)}}. IEEE, \bibinfo{pages}{747--762}.
\newblock


\bibitem[Garcia et~al\mbox{.}(2017)]%
        {garcia2017hey}
\bibfield{author}{\bibinfo{person}{Luis Garcia}, \bibinfo{person}{Ferdinand
  Brasser}, \bibinfo{person}{Mehmet~Hazar Cintuglu},
  \bibinfo{person}{Ahmad-Reza Sadeghi}, \bibinfo{person}{Osama~A Mohammed},
  {and} \bibinfo{person}{Saman~A Zonouz}.} \bibinfo{year}{2017}\natexlab{}.
\newblock \showarticletitle{Hey, My Malware Knows Physics! Attacking PLCs with
  Physical Model Aware Rootkit.}. In \bibinfo{booktitle}{\emph{NDSS}}.
\newblock


\bibitem[Givehchi et~al\mbox{.}(2014)]%
        {givehchi2014control}
\bibfield{author}{\bibinfo{person}{Omid Givehchi}, \bibinfo{person}{Jahanzaib
  Imtiaz}, \bibinfo{person}{Henning Trsek}, {and} \bibinfo{person}{Juergen
  Jasperneite}.} \bibinfo{year}{2014}\natexlab{}.
\newblock \showarticletitle{Control-as-a-service from the cloud: A case study
  for using virtualized PLCs}. In \bibinfo{booktitle}{\emph{2014 10th IEEE
  Workshop on Factory Communication Systems (WFCS 2014)}}. IEEE,
  \bibinfo{pages}{1--4}.
\newblock


\bibitem[Goldschmidt et~al\mbox{.}(2015)]%
        {goldschmidt2015cloud}
\bibfield{author}{\bibinfo{person}{Thomas Goldschmidt},
  \bibinfo{person}{Mahesh~Kumar Murugaiah}, \bibinfo{person}{Christian
  Sonntag}, \bibinfo{person}{Bastian Schlich}, \bibinfo{person}{Sebastian
  Biallas}, {and} \bibinfo{person}{Peter Weber}.}
  \bibinfo{year}{2015}\natexlab{}.
\newblock \showarticletitle{Cloud-based control: A multi-tenant, horizontally
  scalable soft-PLC}. In \bibinfo{booktitle}{\emph{2015 IEEE 8th International
  Conference on Cloud Computing}}. IEEE, \bibinfo{pages}{909--916}.
\newblock


\bibitem[{Gruss} et~al\mbox{.}(2018)]%
        {8418607}
\bibfield{author}{\bibinfo{person}{D. {Gruss}}, \bibinfo{person}{M. {Lipp}},
  \bibinfo{person}{M. {Schwarz}}, \bibinfo{person}{D. {Genkin}},
  \bibinfo{person}{J. {Juffinger}}, \bibinfo{person}{S. {O'Connell}},
  \bibinfo{person}{W. {Schoechl}}, {and} \bibinfo{person}{Y. {Yarom}}.}
  \bibinfo{year}{2018}\natexlab{}.
\newblock \showarticletitle{Another Flip in the Wall of Rowhammer Defenses}. In
  \bibinfo{booktitle}{\emph{2018 IEEE Symposium on Security and Privacy (SP)}}.
  \bibinfo{pages}{245--261}.
\newblock
\urldef\tempurl%
\url{https://doi.org/10.1109/SP.2018.00031}
\showDOI{\tempurl}


\bibitem[Gruss et~al\mbox{.}(2016)]%
        {gruss2016rowhammer}
\bibfield{author}{\bibinfo{person}{Daniel Gruss},
  \bibinfo{person}{Cl{\'e}mentine Maurice}, {and} \bibinfo{person}{Stefan
  Mangard}.} \bibinfo{year}{2016}\natexlab{}.
\newblock \showarticletitle{Rowhammer. js: A remote software-induced fault
  attack in javascript}. In \bibinfo{booktitle}{\emph{International Conference
  on Detection of Intrusions and Malware, and Vulnerability Assessment}}.
  Springer, \bibinfo{pages}{300--321}.
\newblock


\bibitem[Gupta et~al\mbox{.}(2010)]%
        {gupta2010difference}
\bibfield{author}{\bibinfo{person}{Diwaker Gupta}, \bibinfo{person}{Sangmin
  Lee}, \bibinfo{person}{Michael Vrable}, \bibinfo{person}{Stefan Savage},
  \bibinfo{person}{Alex~C Snoeren}, \bibinfo{person}{George Varghese},
  \bibinfo{person}{Geoffrey~M Voelker}, {and} \bibinfo{person}{Amin Vahdat}.}
  \bibinfo{year}{2010}\natexlab{}.
\newblock \showarticletitle{Difference engine: Harnessing memory redundancy in
  virtual machines}.
\newblock \bibinfo{journal}{\emph{Commun. ACM}} \bibinfo{volume}{53},
  \bibinfo{number}{10} (\bibinfo{year}{2010}), \bibinfo{pages}{85--93}.
\newblock


\bibitem[Jia et~al\mbox{.}(2015)]%
        {jia2015coordinate}
\bibfield{author}{\bibinfo{person}{Gangyong Jia}, \bibinfo{person}{Guangjie
  Han}, \bibinfo{person}{Joel~JPC Rodrigues}, \bibinfo{person}{Jaime Lloret},
  {and} \bibinfo{person}{Wei Li}.} \bibinfo{year}{2015}\natexlab{}.
\newblock \showarticletitle{Coordinate memory deduplication and partition for
  improving performance in cloud computing}.
\newblock \bibinfo{journal}{\emph{IEEE Transactions on Cloud Computing}}
  \bibinfo{volume}{7}, \bibinfo{number}{2} (\bibinfo{year}{2015}),
  \bibinfo{pages}{357--368}.
\newblock


\bibitem[Kim et~al\mbox{.}(2014b)]%
        {kim2014architectural}
\bibfield{author}{\bibinfo{person}{Dae-Hyun Kim}, \bibinfo{person}{Prashant~J
  Nair}, {and} \bibinfo{person}{Moinuddin~K Qureshi}.}
  \bibinfo{year}{2014}\natexlab{b}.
\newblock \showarticletitle{Architectural support for mitigating row hammering
  in DRAM memories}.
\newblock \bibinfo{journal}{\emph{IEEE Computer Architecture Letters}}
  \bibinfo{volume}{14}, \bibinfo{number}{1} (\bibinfo{year}{2014}),
  \bibinfo{pages}{9--12}.
\newblock


\bibitem[Kim et~al\mbox{.}(2014a)]%
        {kim2014flipping}
\bibfield{author}{\bibinfo{person}{Yoongu Kim}, \bibinfo{person}{Ross Daly},
  \bibinfo{person}{Jeremie Kim}, \bibinfo{person}{Chris Fallin},
  \bibinfo{person}{Ji~Hye Lee}, \bibinfo{person}{Donghyuk Lee},
  \bibinfo{person}{Chris Wilkerson}, \bibinfo{person}{Konrad Lai}, {and}
  \bibinfo{person}{Onur Mutlu}.} \bibinfo{year}{2014}\natexlab{a}.
\newblock \showarticletitle{Flipping bits in memory without accessing them: An
  experimental study of DRAM disturbance errors}.
\newblock \bibinfo{journal}{\emph{ACM SIGARCH Computer Architecture News}}
  \bibinfo{volume}{42}, \bibinfo{number}{3} (\bibinfo{year}{2014}),
  \bibinfo{pages}{361--372}.
\newblock


\bibitem[Klick et~al\mbox{.}(2015)]%
        {klick2015internet}
\bibfield{author}{\bibinfo{person}{Johannes Klick}, \bibinfo{person}{Stephan
  Lau}, \bibinfo{person}{Daniel Marzin}, \bibinfo{person}{Jan-Ole Malchow},
  {and} \bibinfo{person}{Volker Roth}.} \bibinfo{year}{2015}\natexlab{}.
\newblock \showarticletitle{Internet-facing PLCs-a new back orifice}.
\newblock \bibinfo{journal}{\emph{Blackhat USA}} (\bibinfo{year}{2015}),
  \bibinfo{pages}{22--26}.
\newblock


\bibitem[Kwong et~al\mbox{.}(2020)]%
        {kwong2020rambleed}
\bibfield{author}{\bibinfo{person}{Andrew Kwong}, \bibinfo{person}{Daniel
  Genkin}, \bibinfo{person}{Daniel Gruss}, {and} \bibinfo{person}{Yuval
  Yarom}.} \bibinfo{year}{2020}\natexlab{}.
\newblock \showarticletitle{RAMBleed: Reading bits in memory without accessing
  them}. In \bibinfo{booktitle}{\emph{2020 IEEE Symposium on Security and
  Privacy (SP)}}. IEEE, \bibinfo{pages}{695--711}.
\newblock


\bibitem[Langmann and Rojas-Pe{\~n}a(2016)]%
        {langmann2016plc}
\bibfield{author}{\bibinfo{person}{Reinhard Langmann} {and}
  \bibinfo{person}{Leandro~F Rojas-Pe{\~n}a}.} \bibinfo{year}{2016}\natexlab{}.
\newblock \showarticletitle{A PLC as an Industry 4.0 component}. In
  \bibinfo{booktitle}{\emph{2016 13th International Conference on Remote
  Engineering and Virtual Instrumentation (REV)}}. IEEE,
  \bibinfo{pages}{10--15}.
\newblock


\bibitem[Langmann and Stiller(2019)]%
        {langmann2019plc}
\bibfield{author}{\bibinfo{person}{Reinhard Langmann} {and}
  \bibinfo{person}{Michael Stiller}.} \bibinfo{year}{2019}\natexlab{}.
\newblock \showarticletitle{The PLC as a smart service in industry 4.0
  production systems}.
\newblock \bibinfo{journal}{\emph{Applied Sciences}} \bibinfo{volume}{9},
  \bibinfo{number}{18} (\bibinfo{year}{2019}), \bibinfo{pages}{3815}.
\newblock


\bibitem[Lanteigne(2016)]%
        {lanteigne2016rowhammer}
\bibfield{author}{\bibinfo{person}{Mark Lanteigne}.}
  \bibinfo{year}{2016}\natexlab{}.
\newblock \showarticletitle{How rowhammer could be used to exploit weaknesses
  in computer hardware}.
\newblock \bibinfo{journal}{\emph{SEMICON China}} (\bibinfo{year}{2016}).
\newblock


\bibitem[Lasi et~al\mbox{.}(2014)]%
        {lasi2014industry}
\bibfield{author}{\bibinfo{person}{Heiner Lasi}, \bibinfo{person}{Peter
  Fettke}, \bibinfo{person}{Hans-Georg Kemper}, \bibinfo{person}{Thomas Feld},
  {and} \bibinfo{person}{Michael Hoffmann}.} \bibinfo{year}{2014}\natexlab{}.
\newblock \showarticletitle{Industry 4.0}.
\newblock \bibinfo{journal}{\emph{Business \& information systems engineering}}
  \bibinfo{volume}{6}, \bibinfo{number}{4} (\bibinfo{year}{2014}),
  \bibinfo{pages}{239--242}.
\newblock


\bibitem[Liu et~al\mbox{.}(2011)]%
        {liu2011false}
\bibfield{author}{\bibinfo{person}{Yao Liu}, \bibinfo{person}{Peng Ning}, {and}
  \bibinfo{person}{Michael~K Reiter}.} \bibinfo{year}{2011}\natexlab{}.
\newblock \showarticletitle{False data injection attacks against state
  estimation in electric power grids}.
\newblock \bibinfo{journal}{\emph{ACM Transactions on Information and System
  Security (TISSEC)}} \bibinfo{volume}{14}, \bibinfo{number}{1}
  (\bibinfo{year}{2011}), \bibinfo{pages}{1--33}.
\newblock


\bibitem[McLaughlin and Zonouz(2014)]%
        {mclaughlin2014controller}
\bibfield{author}{\bibinfo{person}{Stephen McLaughlin} {and}
  \bibinfo{person}{Saman Zonouz}.} \bibinfo{year}{2014}\natexlab{}.
\newblock \showarticletitle{Controller-aware false data injection against
  programmable logic controllers}. In \bibinfo{booktitle}{\emph{2014 IEEE
  International Conference on Smart Grid Communications (SmartGridComm)}}.
  IEEE, \bibinfo{pages}{848--853}.
\newblock


\bibitem[Meaney et~al\mbox{.}(2012)]%
        {meaney2012ibm}
\bibfield{author}{\bibinfo{person}{Patrick~J Meaney},
  \bibinfo{person}{Luis~Alfonso Lastras-Monta{\~n}o},
  \bibinfo{person}{Vesselina~K Papazova}, \bibinfo{person}{Eldee Stephens},
  \bibinfo{person}{JS Johnson}, \bibinfo{person}{Luiz~C Alves},
  \bibinfo{person}{James~A O'Connor}, {and} \bibinfo{person}{William~J
  Clarke}.} \bibinfo{year}{2012}\natexlab{}.
\newblock \showarticletitle{IBM zEnterprise redundant array of independent
  memory subsystem}.
\newblock \bibinfo{journal}{\emph{IBM Journal of Research and Development}}
  \bibinfo{volume}{56}, \bibinfo{number}{1.2} (\bibinfo{year}{2012}),
  \bibinfo{pages}{4--1}.
\newblock


\bibitem[Microsoft({[n.\,d.]})]%
        {largepage}
\bibfield{author}{\bibinfo{person}{Microsoft}.}
  \bibinfo{year}{[n.\,d.]}\natexlab{}.
\newblock \bibinfo{booktitle}{\emph{{Large-Page Support}}}.
\newblock
\newblock
\shownote{\url{https://docs.microsoft.com/en-us/windows/win32/memory/large-page-support}}.


\bibitem[Miller et~al\mbox{.}(2013)]%
        {miller2013xlh}
\bibfield{author}{\bibinfo{person}{Konrad Miller}, \bibinfo{person}{Fabian
  Franz}, \bibinfo{person}{Marc Rittinghaus}, \bibinfo{person}{Marius
  Hillenbrand}, {and} \bibinfo{person}{Frank Bellosa}.}
  \bibinfo{year}{2013}\natexlab{}.
\newblock \showarticletitle{XLH: More effective memory deduplication scanners
  through cross-layer hints}. In \bibinfo{booktitle}{\emph{2013 USENIX Annual
  Technical Conference USENIX ATC 13)}}. \bibinfo{pages}{279--290}.
\newblock


\bibitem[Mutlu(2017)]%
        {mutlu2017rowhammer}
\bibfield{author}{\bibinfo{person}{Onur Mutlu}.}
  \bibinfo{year}{2017}\natexlab{}.
\newblock \showarticletitle{The RowHammer problem and other issues we may face
  as memory becomes denser}. In \bibinfo{booktitle}{\emph{Design, Automation \&
  Test in Europe Conference \& Exhibition (DATE), 2017}}. IEEE,
  \bibinfo{pages}{1116--1121}.
\newblock


\bibitem[Oliverio et~al\mbox{.}(2017)]%
        {oliverio2017secure}
\bibfield{author}{\bibinfo{person}{Marco Oliverio}, \bibinfo{person}{Kaveh
  Razavi}, \bibinfo{person}{Herbert Bos}, {and} \bibinfo{person}{Cristiano
  Giuffrida}.} \bibinfo{year}{2017}\natexlab{}.
\newblock \showarticletitle{Secure Page Fusion with VUsion: https://www. vusec.
  net/projects/VUsion}. In \bibinfo{booktitle}{\emph{Proceedings of the 26th
  Symposium on Operating Systems Principles}}. \bibinfo{pages}{531--545}.
\newblock


\bibitem[Petreley(2004)]%
        {petreley2004security}
\bibfield{author}{\bibinfo{person}{Nicholas Petreley}.}
  \bibinfo{year}{2004}\natexlab{}.
\newblock \showarticletitle{Security report: Windows vs linux}.
\newblock \bibinfo{journal}{\emph{The Register}}  \bibinfo{volume}{22}
  (\bibinfo{year}{2004}), \bibinfo{pages}{1--24}.
\newblock


\bibitem[Pietrek(1994)]%
        {pietrek1994peering}
\bibfield{author}{\bibinfo{person}{Matt Pietrek}.}
  \bibinfo{year}{1994}\natexlab{}.
\newblock \showarticletitle{Peering inside the PE: a tour of the win32 (R)
  portable executable file format}.
\newblock \bibinfo{journal}{\emph{Microsoft Systems Journal-US Edition}}
  \bibinfo{volume}{9}, \bibinfo{number}{3} (\bibinfo{year}{1994}),
  \bibinfo{pages}{15--38}.
\newblock


\bibitem[Rakotondravony et~al\mbox{.}(2017)]%
        {rakotondravony2017classifying}
\bibfield{author}{\bibinfo{person}{No{\"e}lle Rakotondravony},
  \bibinfo{person}{Benjamin Taubmann}, \bibinfo{person}{Waseem Mandarawi},
  \bibinfo{person}{Eva Weish{\"a}upl}, \bibinfo{person}{Peng Xu},
  \bibinfo{person}{Bojan Kolosnjaji}, \bibinfo{person}{Mykolai Protsenko},
  \bibinfo{person}{Hermann De~Meer}, {and} \bibinfo{person}{Hans~P Reiser}.}
  \bibinfo{year}{2017}\natexlab{}.
\newblock \showarticletitle{Classifying malware attacks in IaaS cloud
  environments}.
\newblock \bibinfo{journal}{\emph{Journal of Cloud Computing}}
  \bibinfo{volume}{6}, \bibinfo{number}{1} (\bibinfo{year}{2017}),
  \bibinfo{pages}{26}.
\newblock


\bibitem[Razavi et~al\mbox{.}(2016)]%
        {razavi2016flip}
\bibfield{author}{\bibinfo{person}{Kaveh Razavi}, \bibinfo{person}{Ben Gras},
  \bibinfo{person}{Erik Bosman}, \bibinfo{person}{Bart Preneel},
  \bibinfo{person}{Cristiano Giuffrida}, {and} \bibinfo{person}{Herbert Bos}.}
  \bibinfo{year}{2016}\natexlab{}.
\newblock \showarticletitle{Flip feng shui: Hammering a needle in the software
  stack}. In \bibinfo{booktitle}{\emph{25th $\{$USENIX$\}$ Security Symposium
  ($\{$USENIX$\}$ Security 16)}}. \bibinfo{pages}{1--18}.
\newblock


\bibitem[Ruotsalainen(2018)]%
        {ruotsalainen2018hardening}
\bibfield{author}{\bibinfo{person}{Jarno Ruotsalainen}.}
  \bibinfo{year}{2018}\natexlab{}.
\newblock \emph{\bibinfo{title}{Hardening and architecture of an industrial
  control system in a virtualized environment}}.
\newblock \bibinfo{thesistype}{Master's\ thesis}.
\newblock


\bibitem[Sajid et~al\mbox{.}(2016)]%
        {sajid2016cloud}
\bibfield{author}{\bibinfo{person}{Anam Sajid}, \bibinfo{person}{Haider Abbas},
  {and} \bibinfo{person}{Kashif Saleem}.} \bibinfo{year}{2016}\natexlab{}.
\newblock \showarticletitle{Cloud-assisted IoT-based SCADA systems security: A
  review of the state of the art and future challenges}.
\newblock \bibinfo{journal}{\emph{IEEE Access}}  \bibinfo{volume}{4}
  (\bibinfo{year}{2016}), \bibinfo{pages}{1375--1384}.
\newblock


\bibitem[Scholten(2007)]%
        {scholten2007road}
\bibfield{author}{\bibinfo{person}{Bianca Scholten}.}
  \bibinfo{year}{2007}\natexlab{}.
\newblock \bibinfo{booktitle}{\emph{The road to integration: A guide to
  applying the ISA-95 standard in manufacturing}}.
\newblock \bibinfo{publisher}{Isa}.
\newblock


\bibitem[Seaborn and Dullien(2015)]%
        {seaborn2015exploiting}
\bibfield{author}{\bibinfo{person}{Mark Seaborn} {and} \bibinfo{person}{Thomas
  Dullien}.} \bibinfo{year}{2015}\natexlab{}.
\newblock \showarticletitle{Exploiting the DRAM rowhammer bug to gain kernel
  privileges}.
\newblock \bibinfo{journal}{\emph{Black Hat}}  \bibinfo{volume}{15}
  (\bibinfo{year}{2015}), \bibinfo{pages}{71}.
\newblock


\bibitem[Snyder(2014)]%
        {snyder}
\bibfield{author}{\bibinfo{person}{Bill Snyder}.}
  \bibinfo{year}{2014}\natexlab{}.
\newblock \showarticletitle{{Snowden: The NSA planted backdoors in cisco
  products}}.
\newblock \bibinfo{journal}{\emph{InfoWorld}}  \bibinfo{volume}{15}
  (\bibinfo{year}{2014}).
\newblock


\bibitem[Spenneberg et~al\mbox{.}(2016)]%
        {spenneberg2016plc}
\bibfield{author}{\bibinfo{person}{Ralf Spenneberg}, \bibinfo{person}{Maik
  Br{\"u}ggemann}, {and} \bibinfo{person}{Hendrik Schwartke}.}
  \bibinfo{year}{2016}\natexlab{}.
\newblock \showarticletitle{Plc-blaster: A worm living solely in the plc}.
\newblock \bibinfo{journal}{\emph{Black Hat Asia}}  \bibinfo{volume}{16}
  (\bibinfo{year}{2016}), \bibinfo{pages}{1--16}.
\newblock


\bibitem[Stidham(2001)]%
        {stidham2001can}
\bibfield{author}{\bibinfo{person}{Jonathan Stidham}.}
  \bibinfo{year}{2001}\natexlab{}.
\newblock \showarticletitle{{Can hackers turn your lights off: The
  vulnerability of the US power grid to electronic attack}}.
\newblock \bibinfo{journal}{\emph{SANS Institute InfoSec Reading Room}}
  (\bibinfo{year}{2001}).
\newblock


\bibitem[Swierczynski et~al\mbox{.}(2017)]%
        {swierczynski2016interdiction}
\bibfield{author}{\bibinfo{person}{Pawel Swierczynski}, \bibinfo{person}{Marc
  Fyrbiak}, \bibinfo{person}{Philipp Koppe}, \bibinfo{person}{Amir Moradi},
  {and} \bibinfo{person}{Christof Paar}.} \bibinfo{year}{2017}\natexlab{}.
\newblock \showarticletitle{{Interdiction in practice—Hardware Trojan against
  a high-security USB flash drive}}.
\newblock \bibinfo{journal}{\emph{Journal of Cryptographic Engineering}}
  \bibinfo{volume}{7}, \bibinfo{number}{3} (\bibinfo{year}{2017}),
  \bibinfo{pages}{199--211}.
\newblock


\bibitem[Tatar et~al\mbox{.}(2018)]%
        {tatar2018throwhammer}
\bibfield{author}{\bibinfo{person}{Andrei Tatar},
  \bibinfo{person}{Radhesh~Krishnan Konoth}, \bibinfo{person}{Elias
  Athanasopoulos}, \bibinfo{person}{Cristiano Giuffrida},
  \bibinfo{person}{Herbert Bos}, {and} \bibinfo{person}{Kaveh Razavi}.}
  \bibinfo{year}{2018}\natexlab{}.
\newblock \showarticletitle{Throwhammer: Rowhammer attacks over the network and
  defenses}. In \bibinfo{booktitle}{\emph{2018 USENIX Annual Technical
  Conference (USENIX ATC 18)}}. \bibinfo{pages}{213--226}.
\newblock


\bibitem[Tiegelkamp and John(1995)]%
        {tiegelkamp1995iec}
\bibfield{author}{\bibinfo{person}{Michael Tiegelkamp} {and}
  \bibinfo{person}{Karl-Heinz John}.} \bibinfo{year}{1995}\natexlab{}.
\newblock \bibinfo{booktitle}{\emph{IEC 61131-3: Programming industrial
  automation systems}}. Vol.~\bibinfo{volume}{14}.
\newblock \bibinfo{publisher}{Springer}.
\newblock


\bibitem[Van~der Velden(2015)]%
        {macri}
\bibfield{author}{\bibinfo{person}{Lonneke Van~der Velden}.}
  \bibinfo{year}{2015}\natexlab{}.
\newblock \showarticletitle{{Leaky apps and data shots: Technologies of leakage
  and insertion in NSA-surveillance}}.
\newblock \bibinfo{journal}{\emph{Surveillance \& Society}}
  \bibinfo{volume}{13}, \bibinfo{number}{2} (\bibinfo{year}{2015}),
  \bibinfo{pages}{182--196}.
\newblock


\bibitem[Xing and Zhan(2012)]%
        {xing2012virtualization}
\bibfield{author}{\bibinfo{person}{Yuping Xing} {and} \bibinfo{person}{Yongzhao
  Zhan}.} \bibinfo{year}{2012}\natexlab{}.
\newblock \showarticletitle{Virtualization and cloud computing}.
\newblock In \bibinfo{booktitle}{\emph{Future wireless networks and information
  systems}}. \bibinfo{publisher}{Springer}, \bibinfo{pages}{305--312}.
\newblock


\bibitem[Ylmaz et~al\mbox{.}(2018)]%
        {ylmaz2018cyber}
\bibfield{author}{\bibinfo{person}{Ercan~Nurcan Ylmaz},
  \bibinfo{person}{B{\"u}nyamin Ciylan}, \bibinfo{person}{Serkan G{\"o}nen},
  \bibinfo{person}{Erhan Sindiren}, {and} \bibinfo{person}{G{\"o}k{\c{c}}e
  Karacay{\i}lmaz}.} \bibinfo{year}{2018}\natexlab{}.
\newblock \showarticletitle{Cyber security in industrial control systems:
  Analysis of DoS attacks against PLCs and the insider effect}. In
  \bibinfo{booktitle}{\emph{2018 6th International Istanbul Smart Grids and
  Cities Congress and Fair (ICSG)}}. IEEE, \bibinfo{pages}{81--85}.
\newblock


\end{thebibliography}
\vspace{-0.0em}
\section{Appendix}
\label{sec:appendix}

\subsection{\textbf{Automation pyramid}}
\label{subsec:Automation pyramid}

The automation pyramid is a graphical representation of the layers of automation within a typical industry (Fig. \ref{fig:Automation pyramid}). It has five different levels of integrated devices. The name of the five levels and their components are briefly described below :

\vspace{-0.50em}
\begin{figure}[ht!]
\centering
\includegraphics[width=0.3\textwidth,height=0.15\textheight]{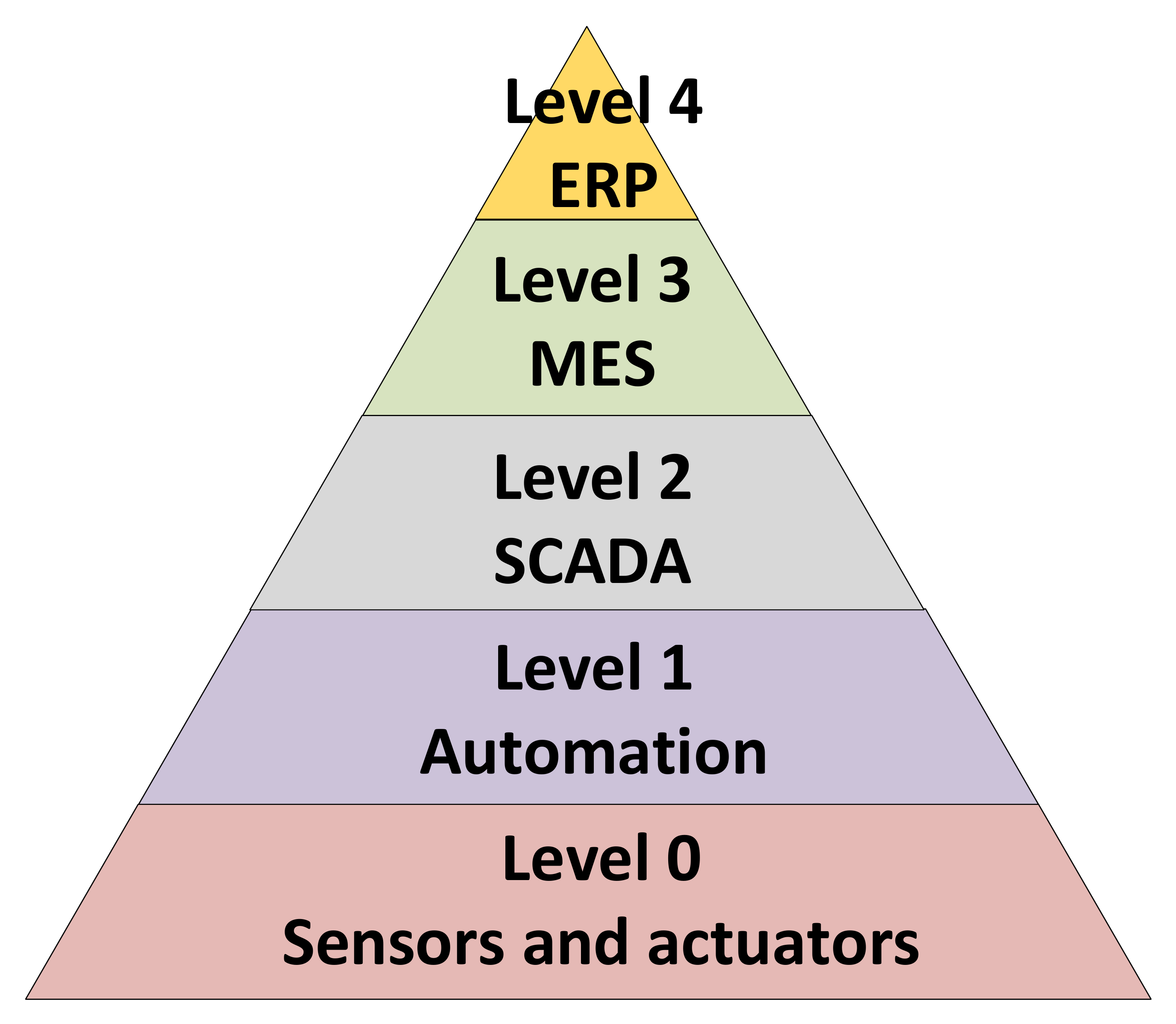}
\vspace{-0.5em}
\caption{Automation pyramid in a typical Industry.}
\label{fig:Automation pyramid}
\vspace{-0.5em}
\end{figure}

\textbf{Level 0 - Sensors and actuators:} This is the bottom level of the automation pyramid and comprises wide variety of sensors and actuators including measurement instruments, communication protocols, and actuators.

\textbf{Level 1 - Automation:} This level is made up with different controllers, such as PLCs, proportional-integral-derivative.

\textbf{Level 2 - SCADA:} This level consists of data acquisition system, human-machine interface, monitoring interfaces, etc.

\textbf{Level 3 - MES:} This level has management execution system (MES) for monitoring the entire process.

\textbf{Level 4 - ERP:} This level is made up with enterprise resource planning (ERP) which is responsible for the integrated management of main business processes.

\subsection{\textbf{PLCs and Industry 4.0}}
\label{subsec:Industry 4.0 and Industrial Control System}

As Programmable Logic Controllers (PLCs) are one of the key ingredients of ICSs, Industry 4.0 drives new approaches in the PLC design \cite{langmann2016plc}. Historically, PLCs were originally designed to support three main concepts, namely programmability, reliability and, real-time response. Different programmable platforms, such as microprocessors, FPGAs, Hard Processor Systems (HPS) are chosen to support programmability in PLCs, as these hardware are programmable in run time in onsite industrial premises following the IEC 61131 key programming standard. Moreover, the standard IEC 61131 is developed in such a way to ensure reliability and real-time response by treating PLCs as logically independent with its own, individual configuration. 

An architecture like this may provide predictable outcomes with a low likelihood of failure, but on the flip-side, it turns out to be progressively lumbering when confronted with developments in IIoTs that require noteworthy adaptability. The IIoTs require the cooperation of individual PLCs on a much deeper level. Moreover, individual PLCs likewise need to work considerably more closely with each other within the industry and remotely, to the web-server and cloud, for instance. 
 
\subsection{\textbf{PLCs interface for basic web technologies}}
\label{subsec:PLCs and basic web technologies}

Todays PLCs have an interface that can be connected to a web-server via a device gateway. The device gateway is integrated into the existing PLC controllers that can support web-compatible protocol required for communication with the IP network. The web-server can connect to the PLC controller using HTML pages that enables a browser-based communication and diagnosis of the PLCs. The web-server can read and write control variables and collect measurement data from PLCs, with restrictions. Sometimes, this web-server is referred to as a ``thin server" having enough computing resources to support local client/server network architecture.

\subsection{\textbf{Implemented protocols}}
\label{subsec:implemented protocols}

Different protocols exist in different layers of ICSs. 
Typically IEC 61158 standard protocols are used in communication between PLCs and  sensors. Here PLCs act as master, and sensors act as slaves. IEC 61158 standard contains a total of nine protocols: Fieldbus, Common Industrial Protocol (CIP), PROFIBUS/PROFINET, P-NET, WorldFIP, INTERBUS, HART, CC-Link,  and SERCOS. These same protocols can be used between PLCs (master) and cloud adapters (slave). RS-232 or RS-485 based Fieldbus has multiple variants. Modbus and DNP3 are two of the most popular variants. They are widely adopted as a de facto standard and has been modified further over the years into several distinct variants. Moreover, Ethernet-based protocols, such as PROFINET, CC-LINK, SERCOS have lower latency than the Fieldbus protocols. Hence, these are preferred over Fieldbus in today's ICSs. 

As already discussed in Section \ref{subsec:Control programming of PLCs}, the program for basic functions and supervisory controls are implemented in clouds or in web-server. These control programs are implemented using service functions in PLC controllers. A standardized protocol named Device Protocol for Web Services (DPWS) enables service-based access to PLC controllers. As mentioned earlier in Section \ref{subsec:PLCs and clouds}, MQTT and AMQP are used to communicate with PLCs from clouds using an IoT gateway.

\subsection{\textbf{Memory deduplication and KVM}}
\label{subsubsec:Memory deduplication}

Memory deduplication or content-based page sharing is a process that combines/merges identical pages in the physical memory into one page. 
When the same/similar operating systems or applications are running in co-located VPSs, lots of redundant pages with same contents are created on the host system. The amount of redundant pages can be as high as 86\% depending on the operating system and workload \cite{chang2011empirical}, and about 50\% of the allocated memory can be saved through memory deduplication \cite{gupta2010difference}. 
Memory deduplication is a feature in Windows 8.1, Windows 10, and Linux distribution. Due to more reliability, high security, stability, and less cost, Linux is more preferable over Windows in ICSs \cite{petreley2004security}. 
That is why here we consider Linux as our implementation platform for memory deduplication, and the idea is similarly applicable to Windows as well. Let us consider that the cloud in our discussion of ICS runs in the Linux platform. To allocate multiple VPSs in the same cloud, Kernel-based Virtual Machine (KVM) has been introduced in the Linux kernel since 2.6.20. 
Memory deduplication is implemented as Kernel Samepage Merging (KSM) in KVM. 
Next, we discuss how KSM 
is used in our attack model to merge the duplicated .bss section.

\subsection{\textbf{Kernel Samepage Merging (KSM)}}
\label{subsubsec:Use of Kernel Samepage Merging}

When a VPS is started, a process named \texttt{qemu-kvm} of the KVM hypervisor allows KSM to merge identical pages in the memory. KSM has a specific daemon named \texttt{ksmd} that periodically scans a specific region of the physical memory of an application. 
The daemon \texttt{ksmd} can be configured in \texttt{sysfs} files in /sys/kernel/mm/ksm location. 
The \texttt{sysfs} files contain different configurable parameters. Among them, we need to mention two parameters: \texttt{pages\_to\_scan}, and \texttt{sleep\_millisec}. 
The parameter \texttt{pages\_to\_scan} defines how many pages to scan before \texttt{ksmd} goes to sleep, and \texttt{sleep\_millisec} defines how much time \texttt{ksmd} daemon sleeps before the next scan. 
If \texttt{sleep\_millisec} = 500, and \texttt{pages\_to\_scan} = 100, then KSM scans roughly 200 pages per second. These numbers depend upon workload and are configured by the cloud provider accordingly. The values of \texttt{sleep\_millisec} and \texttt{pages\_to\_scan} have a significant influence on the \textit{attack time}. This is discussed in Section \ref{subsec:Attack time}.

\subsection{\textbf{KSM data structure}}
\label{subsec:KSM data structure}

The daemon \texttt{ksmd} periodically scans registered address space and looks for pages with similar contents. KSM reduces excessive scanning by sorting the memory pages by their contents into a data structure, and this data structure holds pointers to page locations. Since the contents of the pages may change anytime, KSM uses two data structures in \textit{red-black} tree format, namely unstable tree and stable tree. Moreover, there are three states of each page in the memory: frequently modified state, sharing candidate yet not frequently modified state, and shared/merged state. The page which is frequently modified is not a candidate to be loaded in a stable or unstable tree of  KSM. The page which has similar contents yet not frequently modified (i.e., unchanged for a period of time) is a candidate to be loaded in unstable tree first. The pages in the unstable tree are not write-protected and liable to be corrupted as their contents are modified. The stable tree contains pointers of all shared/merged pages (i.e., ksm pages), and these pages are sorted by their contents in the stable tree. Each page in the stable tree is write-protected. Hence, whenever any process tries to write in the merged/shared page of the stable tree, a private copy of the page corresponding to that particular process is created first and mapped into the page-table-entry (PTE) of that particular process. Then the process writes in that private copy of the page. This is known as copy-on-write (CoW). As CoW involves the creation of a private copy of the shared/merged page of the stable tree first and then writes to that private page, CoW operation is expensive. Therefore, this takes a longer time compared to a write to a regular page. In other words, a longer write time on a page probably indicates that the page is already merged/shared in the stable tree by \texttt{ksmd} daemon. This longer write time in CoW process works as a side channel \cite{bosman2016dedup} and provides an indication that the page is already merged with another page having  similar contents. 

\begin{figure}[ht!]
\vspace{-01.0em}
\centering
\includegraphics[width=0.48\textwidth,height=0.15\textheight]{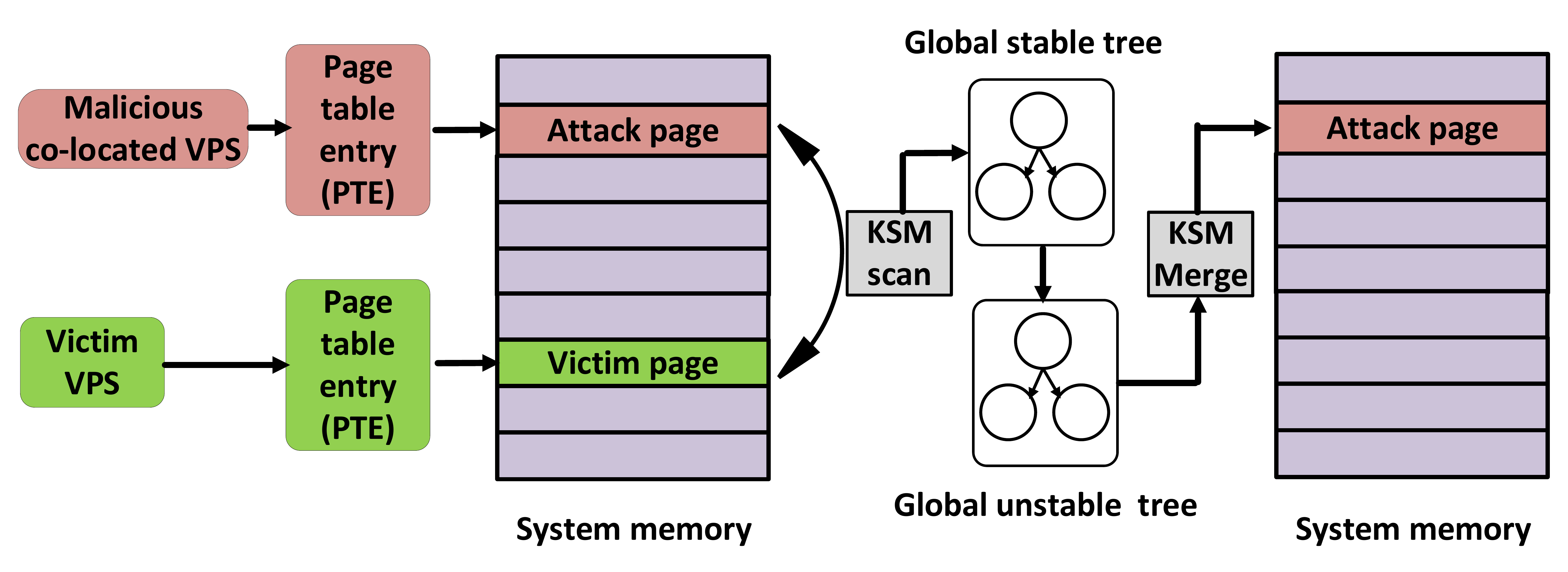}
\vspace{-1.505em}
\caption{Merging \textit{attack page} with the \textit{victim page}.}
\label{fig:ksm merge}
\vspace{-01.00em}
\end{figure}

\section{MemoryDeduplication+Rowhammer}
\label{appendix:sec:Memory deduplication + Rowhammer}

\subsection{\textbf{Process of merging the duplicated .bss section}}
\label{subsubsec:Process of merging the duplicated .bss section}

The process of merging the duplicated .bss section is shown in Fig. \ref{fig:ksm merge}. As discussed earlier, 
the .bss section of the target control DLL is page aligned and is mapped to a page in the physical memory. Let us denote this page as the \textit{victim page}. 
Similarly, the duplicated .bss section of the target control DLL file is also mapped to a different page in the memory. Let us denote this page as the \textit{attack page.} The \textit{attack page} and \textit{victim page} both have same contents. The only difference between them is that the \textit{attack page} is provided by the attacker, whereas the \textit{victim page} is coming from the victim VPS.

The daemon \texttt{ksmd} of the KVM checks the contents of the \textit{attack page} and the \textit{victim page} in the registered address space. Either the \textit{attack page} or the \textit{victim page} is available to the daemon \texttt{ksmd} depending upon their order of arrival in the memory. If the \textit{victim page} arrives first, the daemon \texttt{ksmd} marks this page as a candidate page to be merged. At first, this candidate page is searched in the stable tree using \texttt{memcmp()}. As this candidate page is not still available in the stable tree, it is then searched in the unstable tree by recalculating the checksum over the candidate page. If the checksum has not been changed, the daemon \texttt{ksmd} searches the unstable tree for this candidate page (\texttt{unstable\_tree\_search()}). In this case, as the occurrence of the candidate page (i.e., \textit{victim page}) is first in the unstable tree, this candidate page cannot be found in the unstable tree. As a consequence, a new node is created in the unstable tree for this candidate page (i.e., \textit{victim page}). In the next step, when the \textit{attack page} arrives in the memory, the daemon \texttt{ksmd} marks this page again as the candidate page and searches this page in the unstable tree. As the content of the candidate page (i.e., \textit{attack page}) is same as the \textit{victim page}, this candidate page (i.e., \textit{attack page}) will be merged with the similar node (i.e., \textit{victim page}), which is created in the prior step, in the unstable tree. Then this node of the unstable tree will be merged into the stable tree. If a new candidate page arrives in the memory, this process iterates again.

\subsection{\textbf{Rowhammering on the merged .bss section}}
\label{subsec:Rowhammer attack on merged .bss section}

In Section \ref{subsec:Merging the duplicated .bss section}, we discuss that how the target \textit{victim page} 
is merged with the \textit{attack page} 
using the memory deduplication technique. Note that the attacker cannot simply write to his attack page (i.e., deduplicated page) to  change any data, as simply writing to the deduplicated page by the attacker triggers a CoW (Section \ref{subsubsec:Use of Kernel Samepage Merging}) event to isolate the attack page from the victim page, and the main goal of the KSM may become invalid. That is the reason why the attacker needs something else to corrupt the deduplicated page without triggering the CoW event. Thanks to the Rowhammer bug present in DRAM, Rowhammer can be used to flip bits directly on the DRAM without triggering any CoW event. 

Rowhammer \cite{kim2014flipping} is a widespread vulnerability in recent DRAM devices in which repeatedly accessing a row of DRAM can cause bit flips in adjacent rows. To reliably craft our Rowhammer exploit on the deduplicated page, we have to overcome many challenges. The detail of these challenges is explained as follows.

\vspace{-01.0em}
\begin{figure}[ht!]
\centering
\includegraphics[width=0.48\textwidth,height=0.2\textheight]{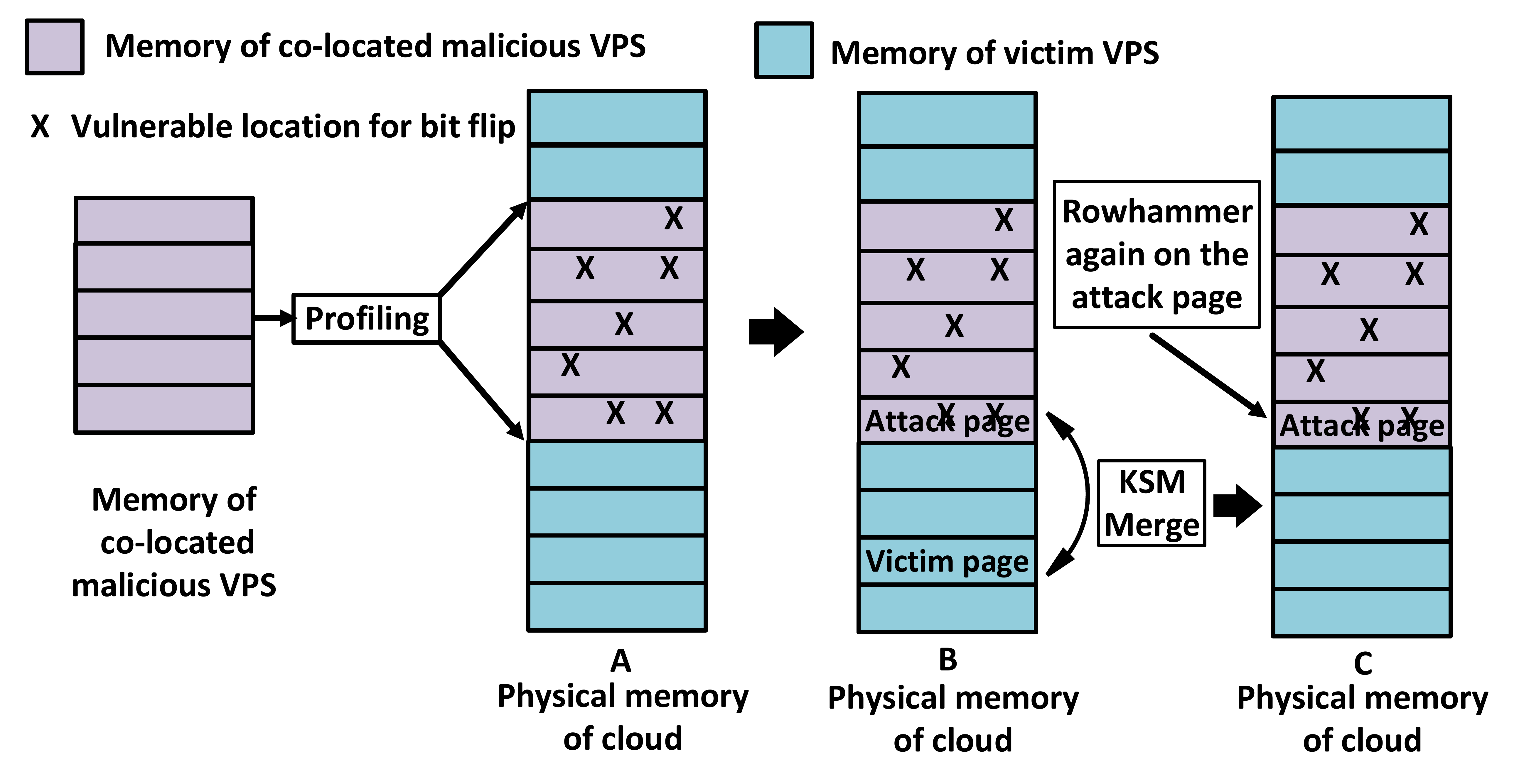}
\vspace{-01.5em}
\caption{(A) Profiling the memory of the cloud. (B) Placing \textit{attack page} in the vulnerable location. (C) After KSM merging, \textit{victim page} is backed by the \textit{attack page}.}
\label{fig:rowhammer}
\vspace{-01.0em}
\end{figure}

\subsection{\textbf{Profiling the vulnerable locations of physical memory}}
\label{subsubsec: Profiling the vulnerable locations of physical memory}

A property of the Rowhammer is that the Rowhammer induced bit-flips tend to be repeatable. A memory location, where a bit flip occurs for the first time, there is a high chance that bit-flips will be reproducible in that location again. Therefore, it is possible to estimate whether a memory location of a DRAM tends to flip. This knowledge of exploitable bit locations is critical for the attacker to successfully exploit the Rowhammer bug from the co-located malicious VPS. 

Therefore, the first step to initiate the Rowhammer attack is to find the \textit{aggressor/victim} addresses in the physical memory of the running system. We name this step as profiling (Fig. \ref{fig:rowhammer}(A)). The \textit{aggressor} addresses are the memory locations within the process's virtual address space that are hammered, and the \textit{victim} addresses are the memory locations where the bit flips occur. For a successful Rowhammer bit flip, the aggressor rows and the victim rows  should be located in different rows but within the same bank of the DRAM chip. If the aggressor rows and the victim rows are located in the different banks of the DRAM chip, the Rowhammer exploit may only read/write from  those bank's \textit{row-buffers} without activating aggressor rows repeatedly. This may not cause any bit-flip in the physical location of the DRAM chip. Therefore, before starting the profiling step, the attacker must ensure that aggressor rows satisfy the \textit{"different rows, same bank''} requirement for the Rowhammer.

\vspace{0.0em}

\subsection{\textbf{Refining the profiling step}}

To ensure different rows but same bank location of the aggressor rows, there are different methods. One method is to use physical addresses of the DRAM rows using an absolute physical address or relative physical address information. The absolute physical address information may not be available by the malicious VPS of the attacker. The relative physical address information can be achieved by using \textit{large pages} \cite{largepage} in Windows VPS. To use the large page support in Windows, the large page option should be activated first in the victim VPS, but it may not be explicitly turned on in the victim VPS. Therefore, double-sided Rowhammering is not a suitable way for the profiling step in the context of ICSs \cite{seaborn2015exploiting}. Another method is to use random address selection. This is a simpler approach, and the attacker does not need to know the absolute physical address or relative physical address of DRAM.  To keep the attack model simpler and easily exploitable, \textit{BayesImposter} uses this random address selection approach for profiling the bit-flippable memory locations of the physical memory. This approach also falls in the category of single-sided Rowhammering.

In the random address selection approach, the attacker allocated a large block of memory of 2 GiB using a large array filled with doubles. A value of $1.7976931348623157 \times 10^{308}$ is stored as double that gives 1 in  memory locations. Next, the attacker randomly picks virtual aggressor addresses from each page of this large memory block and reads $2\times10^6$ times from each random aggressor address of that page. Then  the attacker moves to the next page and repeats the same steps. As the attacker can know the number of banks of the running system from his VPS, he can calculate his chance of hammering addresses in the same bank. For example, in our experimental setup, the machine has 2 Dual Inline Memory Modules (DIMMs) and 8 banks per DIMM. Therefore, the machine has 16 banks, and the attacker has 1/16 chance to hit aggressor rows in the same bank. This 1/16 chance is high for the attacker. Moreover, the attacker hammers 4 aggressor rows in the same iteration that increases the chance of having successful Rowhammering.

After finishing hammering the entire block of memory, the attacker checks the array for possible bit flips. If any bit-flip occurs on any page, the attacker records that page and the offset. In this way, the attacker profiles the memory for vulnerable page/location, where a bit flip is more probable. After profiling, the attacker has aggressor/victim addresses in hand.

The next step is to place the target \textit{victim page} (i.e., page aligned .bss section of the target control DLL) in one of these vulnerable pages. This memory placement  must be done for a successful bit-flip in the target \textit{victim page}. This process is discussed next.

\subsection{\textbf{Placing the target victim page in the vulnerable location}}
\label{subsubsec: Placing the target victim page in the vulnerable location}

As the attacker has aggressor/vulnerable addresses from the profiling step, 
the attacker places the \textit{attack page} in the vulnerable addresses first (Fig. \ref{fig:rowhammer}(B)). 
When the target victim VPS starts, the target victim page 
is merged with the attacker's provided \textit{attack page} using the memory deduplication process (Section \ref{subsec:Merging the duplicated .bss section}). 
Therefore, after merging with the \textit{attack page}, as the \textit{attack page} is used to back the memory of the \textit{victim page}, then, in effect, the attacker controls the physical memory location of the \textit{victim page}. As the \textit{attack page} is placed in the vulnerable addresses for possible bit-flip, then, in effect, the target \textit{victim page} is also placed in the same vulnerable location for possible bit-flip ((Fig. \ref{fig:rowhammer}(C)).

\subsection{\textbf{Rowhammering on the aggressor rows}}
\label{subsubsec:Rowhammering on the aggressor rows}

From the profiling step, the attacker knows the aggressor rows for the vulnerable memory locations. After placing the \textit{attack page} in one of the vulnerable locations, the attacker hammers again on the aggressor rows corresponding to that vulnerable location ((Fig. \ref{fig:rowhammer}(C)). This results in bit-flips in the \textit{attack page} that in effect changes the value of the control programming and supervisory control related variables in the .bss section of the target control DLL.

\end{document}